\documentclass[graybox]{svmult}

\usepackage{type1cm}        
\usepackage{makeidx}         
\usepackage{graphicx}        
\usepackage{multicol}        
\usepackage[bottom]{footmisc}
\usepackage{sidecap}

\usepackage{newtxtext}       %
\usepackage{newtxmath}       

\makeindex             

\usepackage[pdftex,colorlinks=true,bookmarks=false,citecolor=blue,urlcolor=blue]{hyperref} 
\usepackage{float}

\begin{document}

\title*{Terrestrial Laser Interferometers}

\author{Katherine L Dooley, Hartmut Grote and Jo van den Brand}

\institute{Katherine L Dooley \at Cardiff University, Cardiff CF24 3AA United Kingdom, \email{dooleyk@cardiff.ac.uk}
\and Hartmut Grote \at Cardiff University, Cardiff CF24 3AA United Kingdom, \email{groteh@cardiff.ac.uk}
\and Jo van den Brand \at Nikhef and Maastricht University, The Netherlands, \email{jo@nikhef.nl}}


%
%
\maketitle

\abstract*{Please use the 'starred' version of the \texttt{abstract} command for typesetting the text of the online abstracts (cf. source file of this chapter template \texttt{abstract}) and include them with the source files of your manuscript. Use the plain \texttt{abstract} command if the abstract is also to appear in the printed version of the book.}

\abstract{Terrestrial laser interferometers for gravitational-wave detection made the landmark first detection of gravitational waves in 2015. We provide an overview of the history of how these laser interferometers prevailed as the most promising technology in the search for gravitational waves. We describe their working principles and their limitations, and provide examples of some of the most important technologies that enabled their construction. We introduce each of the four large-scale laser interferometer gravitational-wave detectors in operation around the world today and provide a brief outlook for the future of ground-based detectors.}

\section{Introduction: A historical perspective}
\label{sec:1}

Albert Michelson\index{Michelson, Albert} reportedly was a `hard-core' physicist, dedicating pretty much all of his time to research. He was interested early on in improving methods to
measure the speed of light, and to this end he developed the instrument
carrying his name today, the Michelson interferometer.
His invention is of course best known in the history of physics for the
null result testing the ether hypothesis via the attempt to measure differences in the speed of light that travels in different directions. 
While it is disputed to what extent this famous
null result triggered the development of special relativity, it certainly
lent credence to Einstein's theory of 1905. By 1915 Einstein had developed the general theory of relativity, which predicted the existence
of gravitational waves\index{gravitational waves}, though it took decades to convince 
most physicists of the existence and also of the possibility to measure
these waves~\cite{Kennefick:2007} .

The interesting twist here is that much-enhanced successors of Michelson's interferometer first detected gravitational waves in 2015.
These km-scale terrestrial laser interferometers of today are more than ten orders of magnitude more sensitive than the model that Michelson and Morley used for their ether experiment.
In this chapter, we will examine how this astonishing improvement was achieved.

\subsection{Resonant mass detectors}
\begin{figure}[htb]
\sidecaption
\includegraphics[scale=.75]{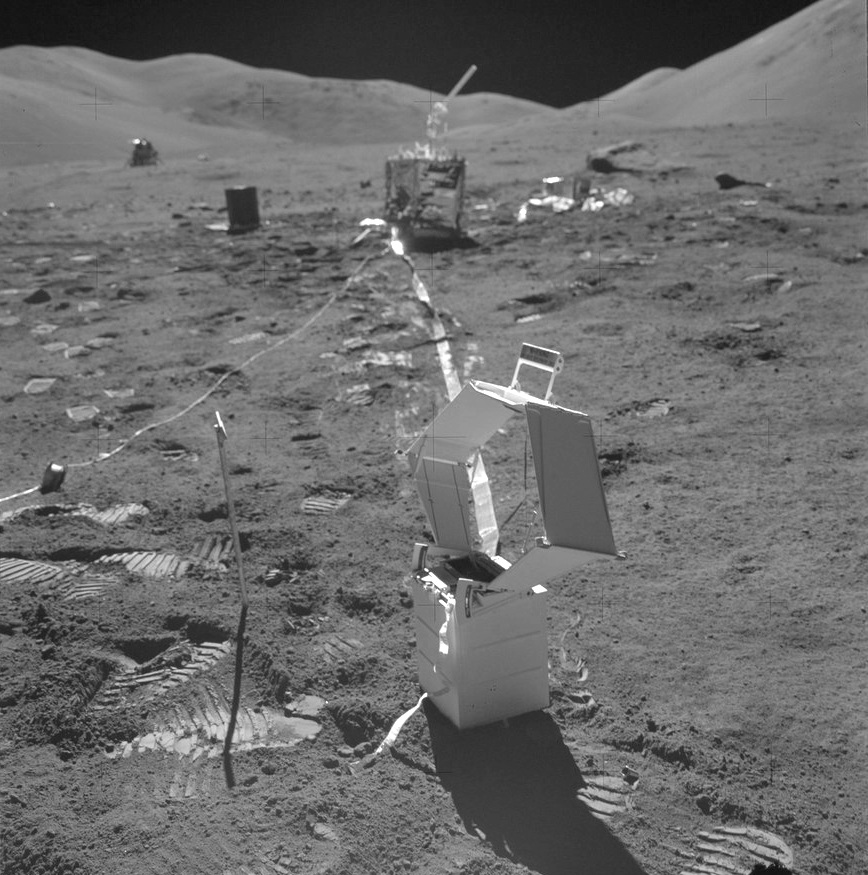}
\caption{A gravimeter on the Moon, which Weber had convinced NASA to deliver on their last Apollo mission. The gravimeter was to measure vertical accelerations of the Moon's surface that could be caused by gravitational waves coupling to the Moon's quadrupolar eigenmode. 
The instrument can be seen in the foreground, with wires running to service stations further back. Photo courtesy: NASA.}
\label{fig:moon}       
\end{figure}

Notwithstanding the title of this chapter, we would like to emphasize here the pioneering work of Joseph Weber, which started the field of experimental gravitational-wave physics. In the late 1950s, Weber contributed to the forming consensus that gravitational waves could indeed be measured, and he set out on a program to attempt the feat with so-called resonant mass detectors. These detectors are massive objects of cylindrical or spherical shape whose mechanical eigenmodes may be excited by passing gravitational waves. Weber claimed to have detected gravitational waves with his detectors in the late 1960s, which spurred several research groups around the globe to attempt replication. In Fig.\,\ref{fig:moon}, we highlight a less well-known episode of Weber's work, where he attempted to use the Moon as a resonant-mass detector ~\cite{Giganti:1977}.

By the mid 70s, no other group had been able to confirm Weber's claims, despite having developed significantly more sensitive detectors. Most scientists today think that Weber was mistaken in the way he analysed his data. Not only was there no confirmation by other groups, but the claimed signal sizes would have meant that most of the mass of the Milky Way would have been converted to energy in the form of gravitational waves. Furthermore, once the sensitivity of laser interferometers had far surpassed that of resonant mass detectors, they also could not confirm the existence of events of the magnitude Weber had claimed he saw.

Once set on this exciting adventure, many research groups did not want to let go of the fascinating prospect of detecting gravitational waves. Subsequently, the experimental community split into two branches. One continued to perfect resonant mass detectors to unprecedented sensitivity levels by cooling ton-scale masses to millikelvin temperatures\cite{Aguiar:2010}; by 2016, however, all of the operating resonant-mass projects had stopped taking data. The other branch that started to develop a new technology would ultimately be successful: laser interferometry.

\subsection{The beginnings of laser interferometry}

The idea of using a Michelson interferometer\index{Michelson interferometer} to measure gravitational waves appears to have surfaced among various scientists independently of one another. According to Joseph Weber's\index{Weber, Joseph} lab notes, this idea came to him soon after the Chapel Hill Conference\index{Chapel Hill conference} in 1957, and he spoke of it in a telephone conversation with his colleague Robert Forward in September 1964~\cite{Collins:2004}. The idea had been published in Russian in a 1962 work by Mikahil Gertsenshtein and Vladislav Pustovoit; see \cite{Gertsenshtein:1963} for an English translation.

Because one could make interferometers long, it was recognized that they had the potential to be sensitive to strains---a length change proportional to distance---of $10^{-21}$ or less, where detections were deemed possible.
But assessing the feasibility of building such an instrument required extensive analyses of noise sources and of the technology that was available.
Experimental physicist Rainer Weiss\index{Weiss, Rainer} carried out much of this early work after he began to think about interferometers\index{interferometer} as gravitational wave detectors in 1969. He calculated how sensitive such an instrument could be and how the influence of various sources of noise could be minimised \cite{Weiss:1972}. Weiss cites the work of Felix Pirani, a British theoretical physicist, and the running of an undergraduate seminar as two of his inspirations.

The theoretical physicist Kip Thorne\index{Thorne, Kip} was interested in gravitational waves early on in his research and was an enthusiastic supporter of Weber. Initially, Thorne was not convinced about developing interferometers for the purpose of gravitational-wave detection. In 1970, in a standard textbook on gravity co-authored with Misner and Wheeler \cite{Misner:2017}, he writes: 
\begin{quotation} Such detectors have such low sensitivity that they are of little experimental interest.\end{quotation} At the time, lasers\index{laser}, in particular, were still very unstable and existing interferometers were far from being sensitive enough to detect gravitational waves. In spite of his skepticism, Thorne maintained contact with Weiss, who considered the use of interferometers as feasible, in principle. 

Robert Forward from Hughes Aircraft Research Laboratory in Malibu, California, and a former member of Weber's team, was the first scientist to begin building an interferometer as a prototype in 1971. With a simple folded arm of effective length of $4.25$\,meters, this instrument achieved about the same sensitivity to gravitational waves as Weber's cylinder\index{Weber cylinder}.\,\cite{Forward:1978} The difference was that the interferometer was sensitive across a broad frequency band, providing a distinct advantage over cylinders which were sensitive in only a narrow band around $1660$\,Hz. The further development of Forward's interferometer was discontinued, however, as he turned his attention to other areas of study.

Beginning in 1972, at the Massachusetts Institute of Technology (MIT) in Cambridge, MA, Weiss\index{Weiss, Rainer} attempted to obtain research funding from the National Science Foundation (NSF)\index{NSF}. It was finally granted in 1975, but initially, Weiss had difficulties getting PhD students to work on his project because it involved lengthy development work. At that time, the resonant mass antennas had been established and the future of interferometers was still uncertain. In 1975, Weiss said in an interview~\cite{Collins:2004}: 
\begin{quotation} We [at MIT] are in a physics department. And ... engineering is not considered respectable physics. To build something and show that it works as predicted, but without making a measurement of anything new does not really count as any achievement.\end{quotation} 
Despite this obstacle, Weiss started with a prototype arm length of $1.5$\,meters and was able to secure funding in 1981 for a study to build a much larger detector with arm lengths in the kilometer range. 

In 1974 in Munich, Germany, a group lead by Heinz Billing turned from resonant mass detectors to interferometers and began to build a laboratory-sized prototype with an arm length of $3$ meters. Using the concept of delay lines\index{delay line}, beams passed through each arm of the interferometer up to $138$ times. This was the world's leading interferometer for many years and served as the prototype for the development and successful demonstration of important new interferometer techniques. This included: hanging the mirrors as pendulums\index{pendulum} by Karl Maischberger to avoid mechanical resonances; the invention of the mode cleaner\index{mode cleaner} by Albrecht R\"udiger and others to suppress laser beam movements; the development of a comprehensive theory of the effect of scattered light\index{scattered light} by Walter Winkler; and the concept of power recycling\index{power recycling}, which was proposed at about the same time by both Roland Schilling in Munich and by Ronald Drever in Glasgow.

In 1983, the construction of a much larger and improved prototype with an arm length of $30$\,meters began on the Garching science campus near Munich. This prototype was the first of its kind in the world to reach shot noise\index{shot noise}, an important limitation to the sensitivity of optical interferometers that had previously only been theoretical. This achievement was to be of decisive importance for the funding of the American LIGO\index{LIGO} project. 

By the end of the 1980s, the peak strain sensitivity of the Garching detector was about $10^{-19}$. This was an improvement of a factor of one-thousand over Weber's cylinders\index{Weber cylinder} of twenty years earlier, in addition to having a much wider bandwidth. The 30-meter prototype\index{30-meter prototype} was in use until 2002 and in its final years it was the first interferometer to demonstrate the combination of power and signal recycling, an optical configuration known as dual recycling\index{dual recycling} (see Sec.\,\ref{subsec:ifo}). It also served as a test facility for the GEO\,600\index{GEO\,600} detector in Germany, prompting the development of numerous techniques such as the ability to keep the suspended mirrors at the correct angle during a measurement. 

In Glasgow, Scotland, beginning in 1975, Drever turned his attention to interferometry, initially studying it in order to achieve a more precise readout from resonant mass antennas. In 1976, Drever began constructing a prototype interferometer with an arm length of $10$\,meters, and applied the concept of Fabry-Perot resonators in the arms. In 1979, following an invitation from Thorne, Drever also led the construction of a 40-meter arm-length prototype at the California Institute of Technology (Caltech) in Pasadena, CA. 
After Drever moved to Caltech permanently in 1983, Jim Hough took over the management of the 10-meter prototype in Glasgow, where Brian Meers would develop the concept of signal recycling\index{signal recycling} ~\cite{Meers:1988}.

In addition to these first significant prototypes, another noteworthy facility is the Australian International Gravitational Observatory (AIGO)\index{AIGO}, located north of Perth. Originally, the construction of an interferometer with arms several kilometres long was planned, but, despite concerted effort, the necessary funds could not be obtained. AIGO is currently a prototype with an arm length of 80 meters and is used for testing high intensity laser power in interferometers \cite{Zhao:2006}. 

In the mid-1980s, after gaining experience with laser interferometer prototypes and developing new techniques, groups in the United States, the United Kingdom and Germany, and later in France and Italy, began applying for research funding for kilometer-sized systems. This was a tall order given that at least \$100\,M would be needed to build instruments that were perceived as having only a small chance of ever measuring gravitational waves. 
Interferometer technology was not yet a mature science and uncertainty remained as to whether or not such large facilities would function sufficiently well. 

In the remainder of the chapter we will introduce some of the principles of laser interferometry and the development of today's terrestrial gravitational-wave detectors. In Sec.~\ref{sec:principles}, we look at the basics of how laser interferometers can be used for gravitational-wave detection, including optical design considerations, relevant noise sources and enabling technologies.
In Sec.~\ref{sec:detectors} we introduce the large-scale terrestrial laser interferometers, with a focus on their individual histories and some
particularities. We conclude with an outlook in Sec.~\ref{sec:outlook}, but also refer the reader to the chapters in this Handbook on research for future detectors and third generation detector technologies.

Many more details on advanced interferometric gravitational-wave detectors can be found in the two-volume book of the same name~\cite{RSG:2019}. A more compact scholarly overview of gravitational wave detectors is provided by Saulson's book~\cite{Saulson:2017} and by a few overview papers~\cite{Adhikari:2013,Bond:2017vy};
an introduction to gravitational-wave detection for a public audience can be found in~\cite{Grote:2019}.

\section{Principles}
\label{sec:principles}
We'll start this section by taking a step back from the assumption that a Michelson interferometer\index{Michelson interferometer} is an appropriate tool to sense gravitational waves and build up the reasoning for this central design choice. 
At their core, terrestrial gravitational-wave detectors are instruments that must be capable of measuring a strain in space-time of the order $10^{-21}$ at frequencies of hundreds to thousands of Hertz. The basic design element of today's detectors makes use of the unique property of light---that its speed is absolute---to probe the distance between two inertial masses that act as markers of space-time coordinates. 
A passing gravitational wave modulates ($\Delta L$) the separation ($L$) of these so-called test masses\index{test mass}, when placed several kilometers apart, typically by less than $10^{-19}$\,m around $100$\,Hz. 

The test masses are mirrors, which provides a means to reflect the laser light dozens of times back and forth. Use of these mirrors as either a delay line\index{delay line} or an optical cavity\index{optical cavity} effectively increases their separation, making any induced space-time strain from gravitational waves ($h=\Delta L/L$) result in all the larger a change in the light travel time.~\footnote{There is a limit to just how big the effective mirror separation should be. After all, if the light experiences both a stretching and shrinking of spacetime, the net length change sensed will be reduced. The optimum situation is therefore when the light samples the mirror separation for exactly one half period of the gravitational wave. Critically, this does mean that not all frequencies of gravitational waves can be simultaneously optimally detected.} 
To ensure the mirrors are protected from external forces other than gravity, a careful consideration of all potential disturbances and a scheme to reduce them is necessary. For terrestrial gravitational wave detectors, the motion of the Earth's surface itself is the most egregious of such disturbing forces. Part of the art of the field of designing and building gravitational wave detectors lies in devising ways to mitigate everything that physically displaces the mirrors as well as everything that prevents achieving the most fundamental precision of  displacement measurements. The result has been the construction of detectors that have pushed the limits of modern technologies, from state-of-the-art seismic isolation systems to low-loss optical coatings to
non-classical light sources and ultra-high-vacuum\index{UHV} systems. 

The Michelson interferometer is often mistakenly assumed as a theoretically required necessity of the design of a gravitational wave detector based on interferometry principles. It is not. In principle, gravitational waves can be measured with a single cavity and a perfect clock. However, a perfect clock does not exist, and thus two sets of two mirrors are used and their respective separations measured simultaneously such that one set can act as a reference for detecting common irregularities of the timing measurement.

Here, the arrangement of these cavities as two perpendicular arms of a Michelson interferometer is critical, with the reason rooted in a particular feature of gravitational waves: that they induce strains in spacetime in a quadrupole configuration (see Ch.\,1 of this Handbook). 
By simultaneously measuring the stretching of one arm and the shrinking of the other, it is assured that sensed length changes resulting from common clock irregularities can be decoupled from the differential effect of gravitational waves. In addition, these two simultaneous length measurements make the response of the detector to gravitational waves up to a factor of two larger compared to a single cavity.
\footnote{The precise response depends on the orientation of the detector with respect to the propagation direction of the gravitational wave.}

In this section we derive how gravitational waves couple to a Michelson interferometer (Sec.\,\ref{subsec:coupling}); we present and motivate the various extensions to a Michelson interferometer design, namely the use of optical cavities (Sec.\,\ref{subsec:ifo}); we discuss the primary noise sources, both fundamental and technical, that limit the sensitivity of the instruments (Sec.\,\ref{subsec:noises}); and we describe the principles of a few select enabling technologies (Sec.\,\ref{subsec:tech}). 
For additional resources summarizing the technologies, techniques and theoretical models used in designing gravitational wave detectors, we direct the reader to \cite{Saulson:2017, Adhikari:2013, Bond:2017vy}.

\subsection{Coupling of gravitational waves to a Michelson interferometer}
\label{subsec:coupling}

Two fundamental ways of looking at how a gravitational wave affects the interferometer are useful to distinguish because of the significant effect they have on how one thinks about the functioning of the detector. 
In a viewpoint which is always valid, gravitational waves change the metric describing the space-time between two freely falling test masses. The coordinates of the test masses and their coordinate separation do not change, although the changing metric does make the proper distance change. If, however, one views the test masses in a proper reference frame, the effect of the gravitational waves is to exert a force on the test masses. Their coordinates \emph{do} change. This viewpoint is only valid for test mass separations that are small compared to the gravitational wave wavelength.

Another duality to ponder for a moment is that of light as both a particle and a wave. A question asked by many a thoughtful student that arises from thinking about the effect of gravitational waves on spacetime is: ``If light waves are stretched by gravitational waves, how can we use light as a ruler to detect gravitational waves?" As before, different reference frames will answer this question differently and ultimately the answer is that the wavelength is irrelevant as explained in \cite{Saulson:1997, Shawhan:2004, Garfinkle:2006}. 

\begin{figure}[ht]
\sidecaption
\includegraphics[width=0.6\textwidth]{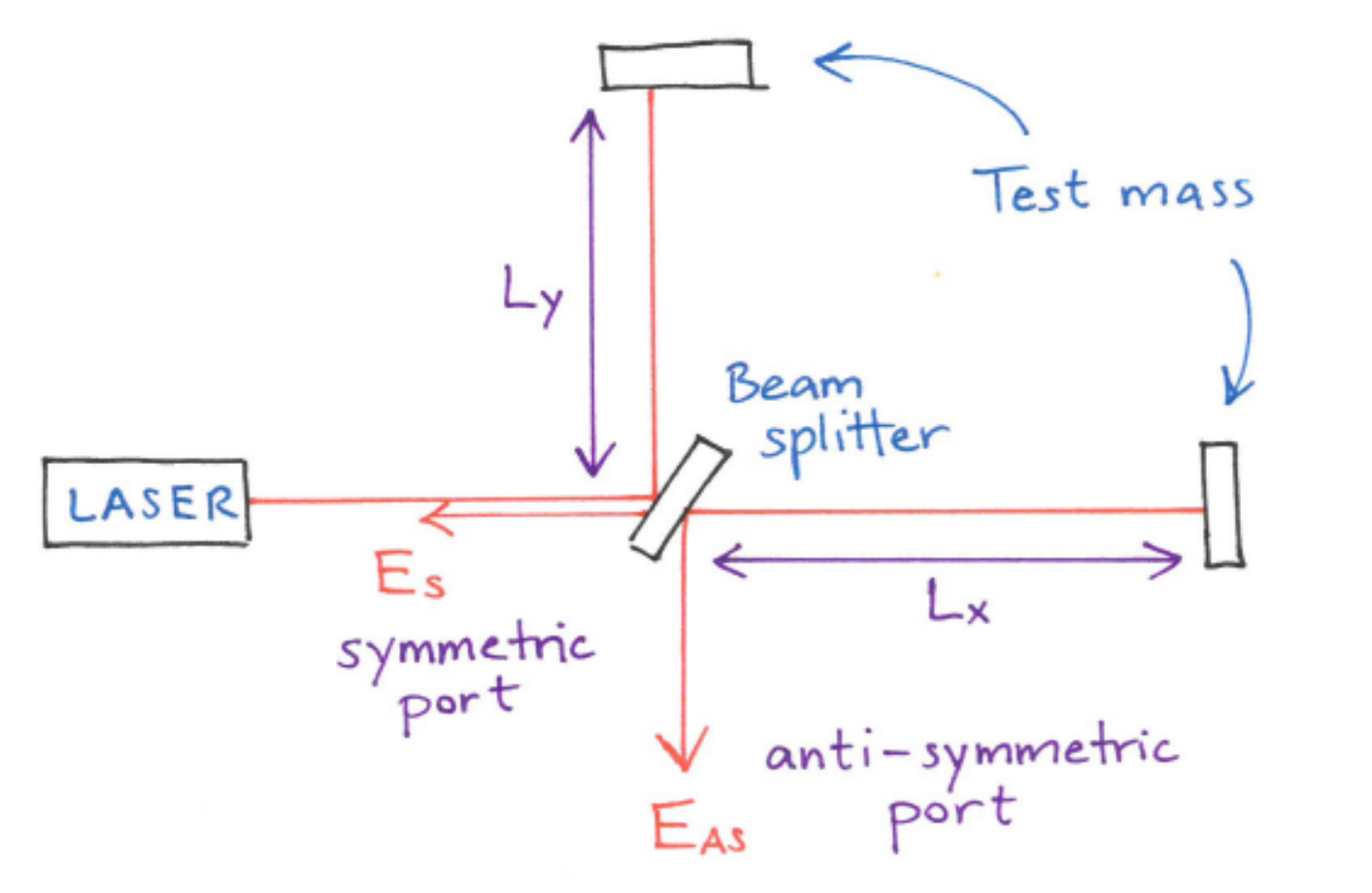}
\caption{A basic Michelson interferometer with arm lengths $L_x$ and $L_y$. Either output port can be used to obtain the gravitational wave signal. A design convention is to use the anti-symmetric port.}
\label{fig:Michelson}
\end{figure}

It becomes clear that a measurable effect exists if we walk through a frame-independent argument of thinking about light as a photon. Consider two wave packets leaving the beam splitter of a basic Michelson interferometer (see Fig.\,\ref{fig:Michelson}) at the same time, each heading down a different arm. If a gravitational wave is present\footnote{It should be noted that $h$ is treated as a constant in Eqs.\,\ref{eq:t+} and \ref{eq:t-}, which assumes that the wavelength of the gravitational wave is much larger than the interferometer arm length. In this case, the temporal variation of $h(t)$ is negligible during the time it takes the photon to make its round trip.} then the amount of time the wave packet takes to make one round trip down a stretched arm and back is
\begin{equation}
    \tau_{\mbox{rt}+} = \frac{2 L}{c} \left( 1+\frac{h}{2} \right).
    \label{eq:t+}
\end{equation}
Likewise, the round-trip travel time for a compressed arm is
\begin{equation}
    \tau_{\mbox{rt}-} = \frac{2 L}{c} \left( 1-\frac{h}{2} \right).
    \label{eq:t-}
\end{equation}
There is a non-zero $2Lh/c$ difference in arrival times at the beam splitter, a quantity one could measure with an accurate clock.

The detector at the beam splitter is not a clock, however, but a photodetector which physically measures the power of the recombined light, and therefore is a proxy for the relative phase of the two returning beams. It is thus informative to express the difference in arrival times as a difference in phase. To do so, we must move away from the photon model and think about the wave model of light where its phase is given by $\phi = \omega \tau$, with $\tau$ the proper time and $\omega$ the angular frequency of the light. Then, the difference in phase between the two light beams after each has completed its round trip is
\begin{equation}
    \Delta \phi_{\mbox{rt}} = \phi_{\mbox{rt}+} - \phi_{\mbox{rt}-} = \frac{2 L}{c} \omega h = 2 k L h
    \label{eq:deltaphi}
\end{equation}
where $k=2\pi/\lambda$ is the wave number.

We can already gain an appreciation for the magnitude of sensitivity required by the gravitational wave detectors. Let's consider the very first gravitational wave detection, GW150914, of a merger of two $\sim 30$\,$M_{\odot}$ black holes $\sim 400$ Mpc away, which produced a strain on Earth of $10^{-21}$ at 100\, Hz about $10$\,ms before their merger \cite{GW150914}. We can use the static strain approximation of Eqs.\,\ref{eq:t+} and \ref{eq:t-} because the wavelength of the gravitational wave is about three orders of magnitude greater than that of the kilometer scale detectors. For simple $4$\,km long Michelson interferometers, the difference in arrival times of wave packet returning from one arm compared to the other is a mere $2.6\times10^{-26}$ seconds.

\subsubsection{The Michelson interferometer response}
\label{sssec:response}
Basic interferometry can be studied using monochromatic, scalar, plane waves as we will do here. A more realistic model must include at least the shape of the beam and additional frequency components, some of which will be discussed later in the chapter. Using the convention of describing an electromagnetic wave by its electric field, and assuming perfectly reflecting end mirrors and a perfect 50/50 beam splitter, one can derive the field at the symmetric and anti-symmetric ports of the interferometer (see Fig.\,\ref{fig:Michelson}):
\begin{eqnarray}
    E_{S} = \frac{E_0}{2}[e^{2 i k L_x}-e^{2 i k L_y}] \\
    \label{eq:E_AS}
    E_{AS} = \frac{E_0}{2}i[e^{2 i k L_x}+e^{2 i k L_y}]
    \label{eq:E_S}
\end{eqnarray}
and easily verify that energy is conserved.

\begin{figure}[ht]
\sidecaption
\includegraphics[width=0.6\textwidth]{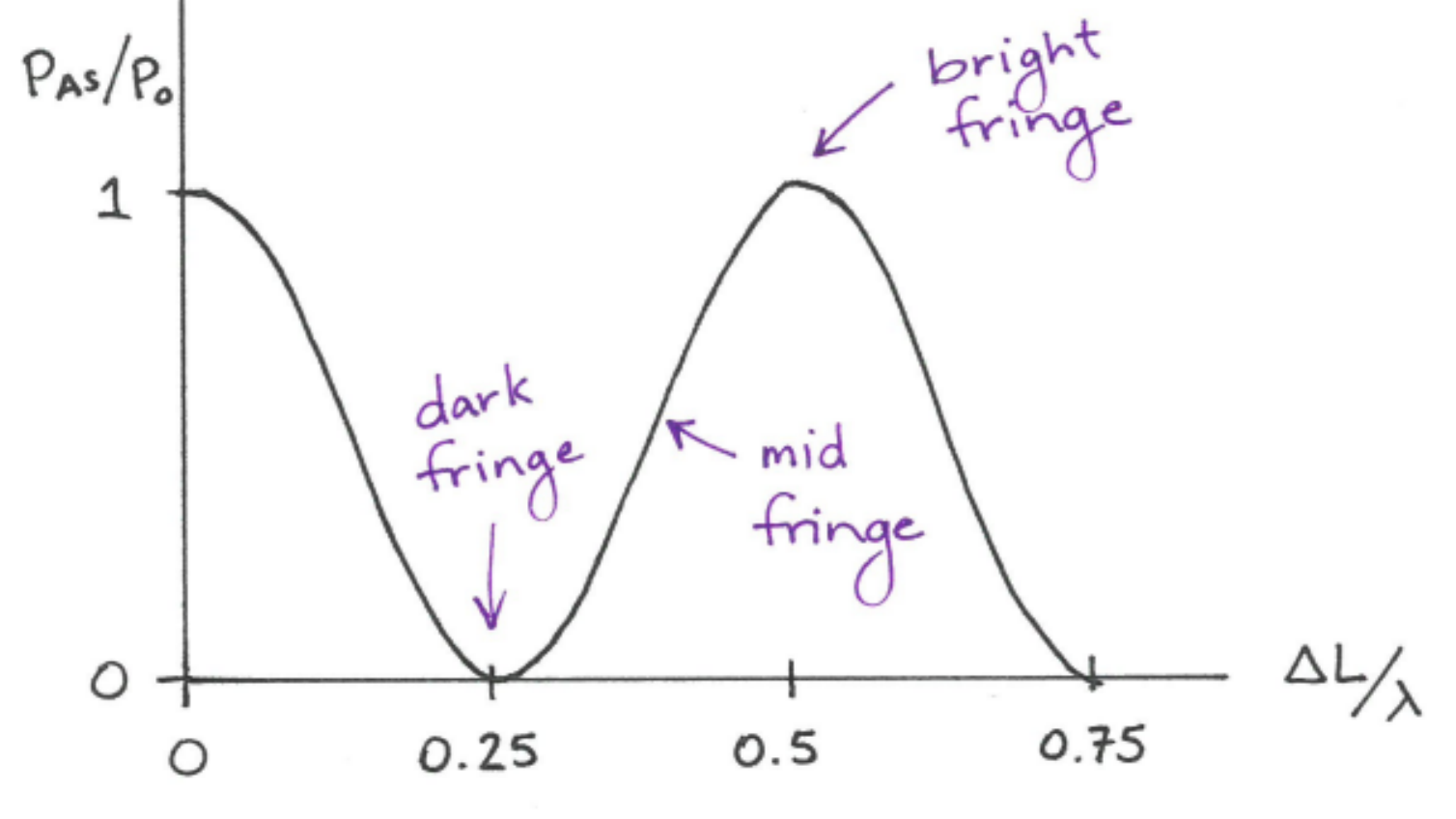}
\caption{The response of a basic Michelson inteferometer to a differential arm length change ($\Delta L$), such as that caused by gravitational waves. Modern gravitational wave detectors are operated at or close to the dark fringe.}
\label{fig:response}
\end{figure}

Because of the quadrupole nature of gravitational waves, their effect on the interferometer is to change the differential arm length, $\Delta L = L_y - L_x$. It is therefore more interesting to express the fields as a function of $\Delta L$ and $\bar{L}=(L_x+L_y)/2$, the common arm length:
\begin{eqnarray}
    E_{S} = E_0 i e^{2 i k \bar{L}} \sin(k \Delta{L}) \\
    E_{AS} = E_0 i e^{2 i k \bar{L}} \cos(k \Delta{L}).
\end{eqnarray}
This response, as measured by a photodetector and normalized by the input laser power $P_0$ is shown in Fig.\,~\ref{fig:response}. We choose to use the anti-symmetric port to extract a signal and we operate the detectors either at or slightly off from the anti-symmetric port dark fringe. The choice of operating point is an important and subtle aspect of interferometer design, though beyond the scope of this chapter so we refer the reader to \cite{RSG:2019, Bond:2017vy}. 
Nonetheless, the basic principle is clear: a modulation of power at the anti-symmetric port can be directly linked to a modulation of the differential arm length.


\subsubsection{Gravitational waves as phase modulation}
\label{sssec:mod}
A different, more comprehensive perspective of describing the effect of gravitational waves on a Michelson interferometer is as a phase modulation\index{phase modulation} of the light in the arms. When a gravitational wave impinges the interferometer, some energy is shifted from one frequency of the laser light, the \emph{carrier}\index{carrier}, to other frequencies, the \emph{sidebands}\index{sidebands}. We must expand our model of the electromagnetic field to no longer be monochromatic, and we must consider non-static gravitational wave strains.

If we extend the model of Eq.\,\ref{eq:deltaphi} to represent non-static gravitational waves, we see that gravitational waves contribute to the total phase picked up by the laser field traversing each arm of the Michelson by the amount $\phi_{GW}(t) = \pm k L h_0 \cos(\Omega_{GW} t)$, where $\Omega_{GW}$ is the angular frequency of the gravitational wave and $h_0$ its amplitude.\footnote{Without loss of generality, we treat here the gravitational wave as monochromatic.} The electric field thus experiences a phase modulation of $e^{i\phi_{GW}}$, which can be expanded using Bessel functions because $h_0\ll1$:
\begin{equation}
    e^{i\phi_{GW}} = 1 \pm i\frac{kLh_0}{2} \exp{(-i\Omega_{GW}t)} \pm i\frac{kLh_0}{2} \exp{(+i\Omega_{GW}t)} \pm ...
    \label{eq:sb}
\end{equation}

The first, zero-th order term represents the so-called carrier field which oscillates at the laser frequency and the other two terms are the lower and upper sideband pair of order one. Gravitational waves thus create new electric fields in the Michelson arms with amplitudes proportional to both the carrier field and $k L h_0 / 2$ and with frequencies shifted away from the carrier frequency by $\pm k \Omega_{GW}$.\footnote{Expansion to the second order is necessary to see that power is indeed transferred from the carrier to the sideband fields.} For stretched arms, the phase of the sideband fields are rotated $+90\deg$ with respect to the carrier field and for compressed arms, they are rotated $-90\deg$. When combined at the beam splitter, the two sets of gravitational wave sidebands create a field at the anti-symmetric port that oscillates at a frequency of $\Omega_{GW}$ with respect to the carrier and has an amplitude proportional to $h_0$.

This signal is too small to be useful on its own. For the toy example presented above of the effect of GW150914 on a simple Michelson, the amplitude of the gravitational wave sidebands is only $10^{-11}$ that of the carrier. If a $200$\,W laser were used, these sidebands\index{sidebands} would produce less than one photon per second at the anti-symmetric port. For gravitational waves like GW150914 that generate strains of $10^{-21}$ for only 10\,ms, less than one out of every $100$ passing gravitational waves would actually produce a photon! 

Solutions for how to measure and amplify these signals include the introduction of local oscillator fields and Fabry-Perot cavities, respectively. The concept of increasing the signal with the use of cavities is addressed in the next section (Sec.\,\ref{subsec:ifo}). Here, we introduce the technical trick of using a local oscillator\index{local oscillator} to measure small signals.
By adding a large field, $E_{LO}$, to the signal, $E_{GW}$, at the anti-symmetric port, the power measured by a photodetector becomes:
\begin{equation}
    P_{AS}=E_{LO}^2 + 2E_{LO}E_{GW} + E_{GW}^2.
\end{equation}
The first term is large and static and the third term (the pure gravitational-wave sidebands) oscillates at $2\Omega_{GW}$, but is proportional to $h_0^2$ and thus negligibly small, as seen in our example above. The middle term is where the benefit of the local oscillator field is relevant: it's a strong signal due to the local oscillator, yet is proportional to $h_0$ and oscillates at $\Omega_{GW}$. Re-calculating our toy example for a local oscillator field of $\sqrt{10\,\rm mW}$ we now obtain of order $10^{10}$ photons per second for the gravitational-wave signal.\footnote{The size of the local oscillator field is determined by technical considerations and, once large enough to dominate other noise sources, does not affect the maximal signal-to-noise ratio that can be achieved.} 
The decision of what field to use as a local oscillator is intricately connected to several technical considerations. Current detectors use a small offset to the dark fringe\index{dark fringe} operating point. In the past, radio frequency sidebands have been used, and in the near future local oscillator\index{local oscillator} fields split off from the carrier light in the interferometer may be used.

\subsection{Extensions to the Michelson Interferometer}
\label{subsec:ifo}

The typical optical configuration used in today's generation of gravitational wave detectors is an extension of the basic Michelson interferometer. The most prevalent additional feature is that of optical cavities, which are included at nearly every port, either for fundamental or technical reasons. Optical cavities can serve several functions that range from creating an effectively longer interferometer arm, to increasing the laser power, to filtering out unwanted spatial modes, laser frequencies and even polarisations of light. 

\begin{figure}[ht]
\sidecaption
\includegraphics[width=0.6\textwidth]{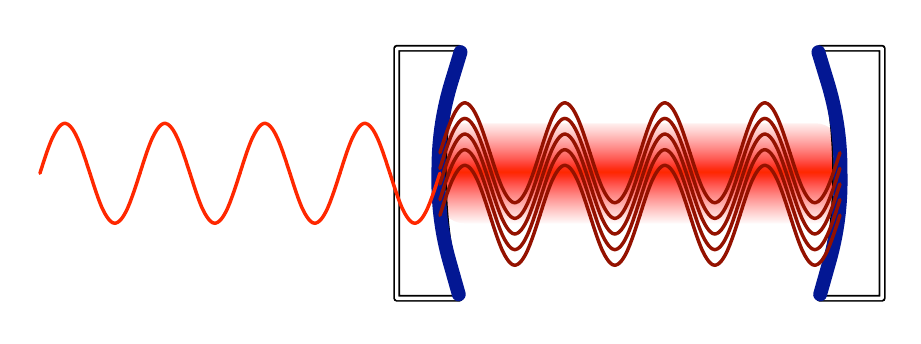}
\caption{A Fabry-Perot cavity\index{Fabry-Perot cavity}. On resonance with the incoming light, the light field inside the cavity is resonantly enhanced by multiple reflections. The reflectivity coefficients of the mirrors set the enhancement level. Image courtesy: Rob Ward.}
\label{fig:cavity}
\end{figure}

Figure\,\ref{fig:cavity} shows an optical cavity consisting of two mirrors, known as a Fabry-Perot cavity\index{Fabry-Perot cavity}, with a light field entering and leaving the cavity on the left hand side.
Assuming monochromatic light of a precise wavelength, the light field is resonantly enhanced between the two mirrors if the round-trip length of the cavity is a multiple integer of the wavelength. This resonant enhancement within a cavity is typically used to increase the precision of an interferometric phase measurement. When the cavity is instead used for spatial and temporal mode-filtering, the reflectivities of the two mirrors are made equal, which results in the filtered light being transmitted to the right hand side.
For an excellent original source on the basics of Fabry-Perot cavities\index{Fabry Perot cavity}, see \cite{Siegman:1986}. 

In considering extensions to the basic Michelson, we assume the interferometer is operated at or close to the dark fringe, such that nearly all carrier power leaves towards the symmetric port and all signal towards the anti-symmetric port\index{anti-symmetric port}. There are two particular extensions to the Michelson interferometer that serve to enhance the gravitational wave signal in two independent ways: the power and signal recycling mirrors. A third extension, Fabry-Perot cavities in the arms, combines the effects of power and signal recycling, and, in combination with power recycling, also accommodates constraints arising from imperfections of the mirrors and the beam splitter.

The power recycling mirror\index{power recycling} is located at the interferometer's symmetric port and forms an optical cavity with each of the Michelson arms. By reflecting back into the interferometer all of the light leaving in the direction of the symmetric port, the laser power in the interferometer can be resonantly enhanced. This is important because as we saw in Eq.\,\ref{eq:sb}, the amplitude of the sidebands\index{sidebands} generated by the gravitational waves is proportional to the amplitude of the carrier field in the Michelson arms. One may question, however, where all of the laser power ultimately goes, since after all, the power build-up in the interferometer does not approach infinite levels even though the laser is always on. It is the reality that the mirrors are not perfect---they scatter and absorb light---that means there is a finite maximum circulating power. To reach that maximum power, the transmissivity of the power recycling mirror is designed to match the arm losses and create a nearly critically coupled (impedance-matched) power-recycling cavity. Today's large interferometers achieve resonant power enhancement factors of about 5000, resulting in hundreds of kW of light power in the arms for input powers of order 100\,W.

The counterpart to the power recycling mirror is the signal recycling\index{signal recycling} mirror, which is located at the interferometer's anti-symmetric port and which forms cavities with each of the Michelson arms. The signal recycling mirror sends the gravitational wave sidebands\index{sidebands} 
back into the interferometer such that they can constructively interfere with the new sidebands being created. The reflectivity of the signal recycling mirror determines how long (on average) the sidebands stay in the interferometer and therefore determines whether the sideband amplitude gets maximally resonantly enhanced or ultimately averaged away or something in between. The signal recycling cavity can therefore be designed to enhance the gravitational wave signals at the frequencies of greatest scientific interest, at the cost of less signal at other frequencies. 
This is done by carefully choosing the reflectivity of the signal recycling mirror.
Power- and signal recycling can be combined, a configuration called dual recycling\index{dual recycling}~\cite{Meers:1988}.

A similar but different extension to the basic Michelson interferometer is the creation of Fabry-Perot cavities\index{Fabry-Perot cavity} in the arms. 
By combining power recycling with arm cavities, the power in the interferometer can be concentrated in the arms and away from the beam splitter. This is important because the beam splitter substrate is transversed by only one of the two beams prior to their interference, making the beam splitter a dominant source of asymmetric optical loss in the interferometer. Any asymmetric loss disrupts the ideal condition that all carrier light leaves towards the symmetric port and none towards the anti-symmetric port. \footnote{The most important reason to maintain perfect decoupling of the two ports is that any carrier light that leaks to the anti-symmetric port adds to the shot noise of the measurement, and thus decreases the signal-to-noise ratio. A secondary reason is that any carrier power heading to the anti-symmetric port means all the less that goes to the symmetric port for recycling.} Power recycling alone can thus typically not achieve the high power levels in the interferometer that are possible when arm cavities are included. In this case, the reflectivity of the power recycling mirror can be adjusted to match the total losses of each arm, ensuring most of the laser power gets dumped into the losses of the arms, which are set to be as equal as possible through careful selection of the available optics.

Traditionally, one often thinks of arm cavities as an effective way to lengthen the arms. This view is equivalent to the effect of signal recycling, in that the storage time of the gravitational wave sidebands is increased.
If such arm cavities are used on their own, they combine the effects of power and signal recycling, but the two parameters cannot be adjusted independently.
When using power recycling and arm cavities, the distribution of power in the interferometer is coupled to the frequency response of the interferometer by choice of the reflectivity of the arm cavity mirrors.
However, when arm cavities and signal recycling are combined, they allow the distribution of power in the interferometer \emph{and} the frequency response to be independently tuned, where the finesse\index{finesse} of the arm cavities determines how power is distributed between the power recycling cavity and the arms. This results in a particular bandwidth of the arm cavities, which can then be modified by the signal recycling mirror.

\begin{figure}[bht]
\centering
\includegraphics[width=0.7\textwidth]{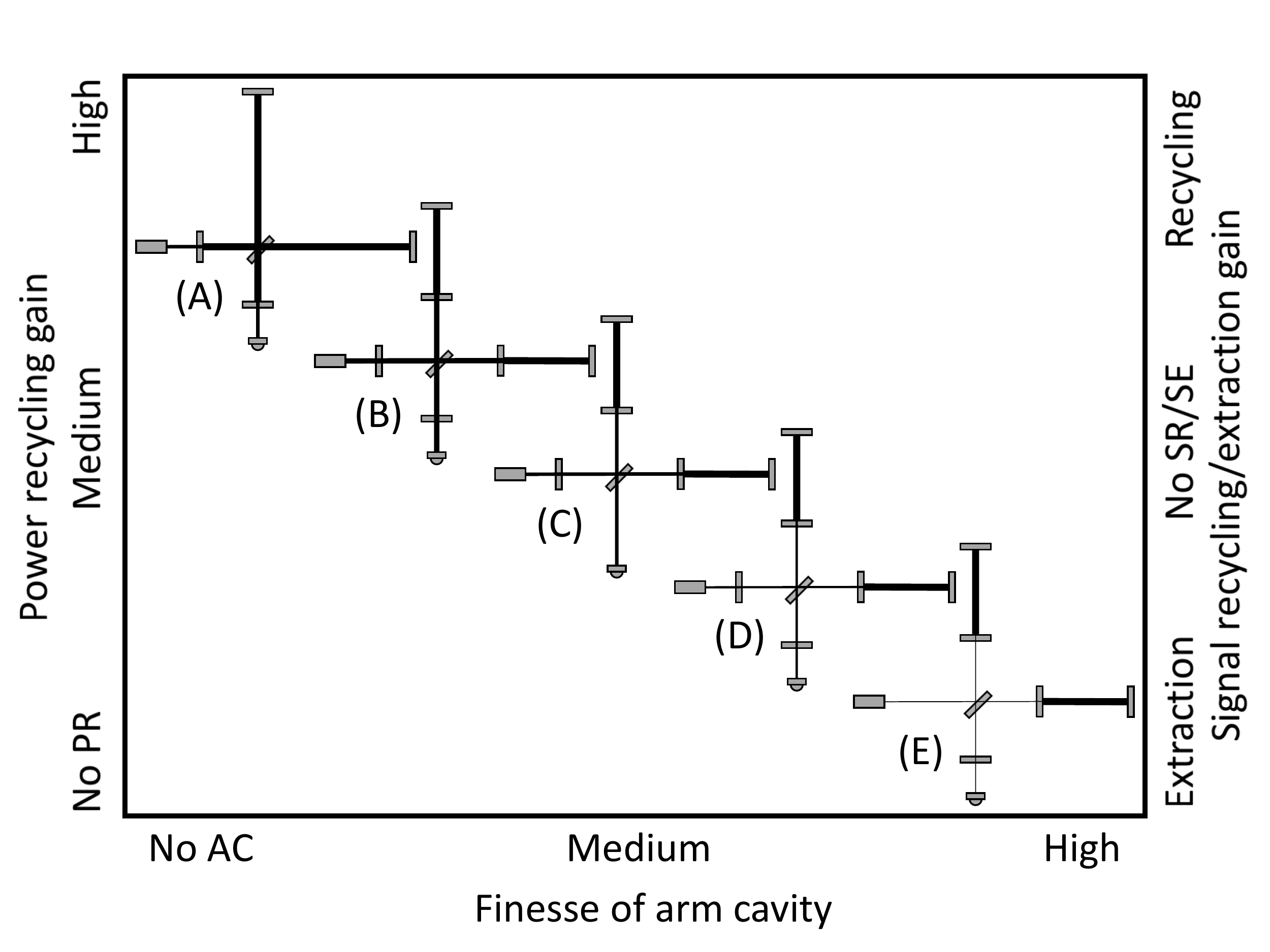}
\caption{A diagram adopted from Seiji Kawamura \cite{RSG:2019} which organizes the concepts of the power and signal recycling cavities in relation to the finesse of the arm cavities. The configurations that have been used by various gravitational wave detectors are as follows: (A) GEO\,600; (C) Initial LIGO and Initial Virgo; (D) Advanced LIGO, Advanced Virgo and KAGRA. In (B) signal recycling is added to the use of arm cavities in a way that further enhances the signal sidebands at low frequencies. 
Option (C) describes the case where the frequency response of the arm cavities is precisely as desired such that no further shaping with a signal recycling mirror is required.
In (D) the signal recycling is operated in RSE mode, decreasing the arm cavity finesse for the signal sidebands, but not for the carrier light. Option (E) describes the extreme where all power buildup is achieved by the arm cavities. Note that in the ideal case, the power in the arms is the same in all options, though in practice this is difficult to achieve in (A). Figure credit: Advanced Interferometric Gravitational-wave Detectors. Reitze, Saulson and Grote, editors. Copyright \copyright 2019 by World Scientific Publishing Co. Pte. Ltd.}
\label{fig:Kawamura}
\end{figure}

Figure~\ref{fig:Kawamura} organizes the parameters characterizing these three types of optical cavities into a single diagram. Here we see how increasing the arm cavity finesse keeps power away from the beam splitter. But due to an increase of losses in the arms with higher finesse, the power recycling gain (the power enhancement factor due to the power recycling cavity) must simultaneously be lowered in order to maintain impedance matching of the power recycling mirror to the arms. 
If arm cavities are present, the modification of bandwidth by the signal recycling mirror can happen in two different ways, either as resonant enhancement of signal sidebands, as in the case of signal recycling without arm cavities, or as resonant de-enhancement, also called resonant sideband extraction (RSE)\index{resonant sideband extraction}. In RSE mode the signal recycling mirror (or RSE mirror, as it may be called in this case) serves to effectively lower the arm cavity finesse, but only for the signal sidebands, not for the carrier field.

\subsection{Noise sources and noise reduction strategies}
\label{subsec:noises}

As outlined above, the interferometers exhibit a particular frequency-dependent response to an incident gravitational wave, which, in short, we refer to as a signal.
As in any measurement, the sought-after signal has to compete with noise sources inherent to the particular measurement undertaken and we therefore
want to maximise the signal-to-noise ratio of the measurement.
While optical cavities\index{optical cavity} can be used to maximise the signal, we turn our attention here to the noise sources and the techniques we use to minimize them.

The sources of noise that contaminate the detector's output may be loosely grouped into two categories: displacement noise\index{displacement noise} and sensing noise\index{sensing noise}. Displacement noises are those that create real motion of mirror surfaces. At very low frequencies seismic noise and the related gravity gradient noise are the dominant displacement noises. In the low- to mid-frequency range thermal motion of the dielectric coatings of the mirrors and of their suspensions  dominate, as does quantum radiation pressure noise. Sensing noises are those that arise during the process of measuring the positions of the mirrors. The primary sensing noise is shot noise, which, in a semi-classical picture, arises from the Poisson statistics of photon arrival time at the photodetector.

\begin{figure}[ht]
\centering
\includegraphics[width=0.8\textwidth]{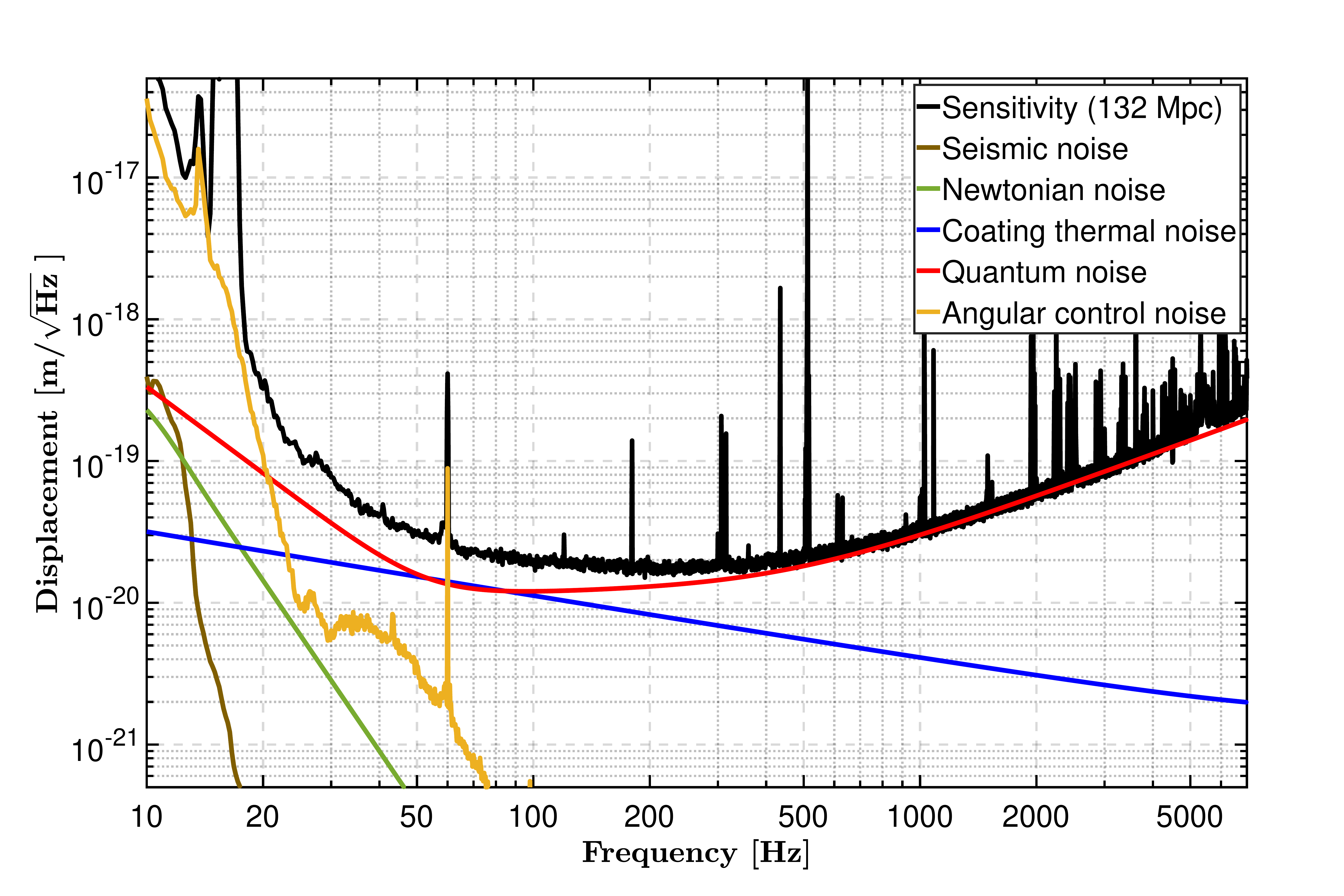}
\caption{A selection of the most critical fundamental displacement and sensing noises, as well as an example technical noise, as shown for the LIGO Livingston detector. Quantum noise and coating thermal noise play the dominant roles in limiting the sensitivity of each of the detectors across most of the frequency band. Angular control noise, a technical noise source, limits the sensitivity below 20\,Hz. The narrow lines of the measured sensitivity represent the thermally-excited violin modes of the test mass suspension fibers. The range is the distance to which a binary neutron star system could be detected (averaged over all sky orientations). Data is from April 2019, during the O3 observing run \cite{LIGOO3Sensitivity}.}
\label{fig:NB_basic}
\end{figure}

Figure~\ref{fig:NB_basic} depicts a typical noise plot showing the inverse of the signal-to-noise ratio, which may be calibrated to displacement or strain. Here we highlight the contributions of some of the most fundamental noise sources for a modern gravitational wave detector in comparison to the total measured displacement noise. The noises are treated as independent of one another and thus are added in quadrature when modeling their total contribution. One will note that the fundamental noises do not fully account for the total measured noise. This is because there are also technical noise sources, such as the angular control noise highlighted in Fig.~\ref{fig:NB_basic}. The subsections that follow describe each of the primary types of noise sources, both fundamental and technical.

\subsubsection{Quantum noise}
\label{sssec:quantum}
\index{quantum noise, shot noise, radiation pressure noise}

Quantum noise is the primary limiting noise source for most of the
sensitive band of gravitational wave detectors. At frequencies above about $50$\,Hz the dominant form of quantum noise is shot noise, the manifestation of Poisson statistics for the counting of detected photons. Shot noise is thus a sensing noise and it limits all position measurements that use classical (i.e. not quantum-manipulated) light.
At frequencies below about $50$\,Hz, quantum noise manifests as radiation pressure noise, where momentum transfer from all photons recoiling
from the freely suspended test masses gives rise to position noise. 

Both shot noise and radiation pressure noise scale with $\sqrt{n}$ for the number of photons $n$. 
In order to increase the signal to shot-noise \emph{ratio}, increasing laser power in the interferometer is a viable concept because the gravitational wave signal is proportional to $n$,
and thus increases more than the shot noise, which is proportional to $\sqrt{n}$.
An increase in laser power, however, leads to an increase in radiation pressure noise at low frequencies, which acts as pure displacement noise.
To keep the radiation pressure effect at bay, one can increase the inertial mass of the test masses up to limits imposed by the materials and fabrication techniques.
Given a fixed test mass, there is then an optimum power to use for any given gravitational-wave target frequency,
since shot- and radiation pressure noise 
are inversely linked by the laser power in the interferometer.

At a deeper level, both shot- and radiation pressure noise can be understood
as the result of the vacuum fluctuations of the electromagnetic field entering the
interferometer from the output port. They are therefore subsumed under the common label of quantum noise. In this description the randomly fluctuating vacuum
field beats with the laser carrier field at the output upon detection to manifest as shot noise,
and at the test mass to manifest as radiation pressure noise.
In pointing this out in a classic paper~\cite{Caves:1981}, Carlton Caves also realised that
in reducing the vacuum fluctuations entering the interferometer, one can reduce one of these noises at the expense of the other.
This can be realised through the use of squeezed vacuum states of light\index{squeezed vacuum},
which have been implemented in GEO\,600 since 2010 \cite{Grote:2013}, and as a permanent addition in LIGO and Virgo since 2019~\cite{LIGOSqz, VirgoSqz}.

The two quadratures of the electromagnetic vacuum field, which can be labeled as amplitude and phase,
are not independent and their product has to obey Heisenberg's uncertainty principle.
Since the amplitude and phase uncertainties relate to radiation pressure and shot noise, respectively, they are relevant at different frequency regions. By reflecting the squeezed beam
off a suitable optical resonator, one can achieve a low noise level in the
phase quadrature for high frequencies and in the amplitude quadrature at low frequencies.
As a result, these so-called filter cavities enable a reduction of quantum noise across the entire
frequency band of a gravitational-wave detector~\cite{FilterCavity}.
For a deeper consideration of quantum noise, the reader is referred to the dedicated chapter within this Handbook.

\subsubsection{Thermal noise}
\label{sssec:thermal}
\index{thermal noise}

Thermal noise\index{thermal cavity} in terrestrial laser interferometers is particularly important in the middle of the sensitivity band for terrestrial gravitational wave detectors, and has been subject to decades of research. It is relevant not only to gravitational-wave detectors, but also to other precision metrology devices where cavity-stabilised lasers are exploited. However, the requirements of gravitational-wave detectors have pushed the understanding of the theory and the development of thermal noise mitigation techniques.

In general, thermally driven fluctuations of mechanical systems can be derived from the application of the Fluctuation-Dissipation (F-D) theorem~\cite{Callen:1951}.
The F-D theorem\index{Fluctuation-Dissipation theorem} provides a link between the fluctuating motion of a system in equilibrium and the dissipative (real) part of the admittance (the inverse of the impedance). 

An early approach to the estimation of thermal noise in mechanical systems of gravitational-wave detectors was to consider them in terms of their vibrational modes, treating each mode as a simple damped harmonic oscillator \cite{Saulson:1990}.
For a terrestrial laser interferometer, it is the thermal noise affecting the surface positions of the test masses that matters most, critical for the measurement of gravitational waves. In this case, surface position fluctuations arise from the internal vibrational modes of the mirrors, the vibrational modes of the fibers suspending the mirrors (also called violin modes), and the pendulum motion of the suspended mirrors. The uncorrelated sum of noises associated with each of these degrees of freedom constitutes the total thermal noise.

It was later understood, however, that the approach of independently treating and adding the noises of all degrees of freedom assumes spatially uniform dissipation mechanisms.
In particular, for the optically coated mirrors of a gravitational wave detector, this condition proved to be too much of an idealization, such that other methods had to be found in order to estimate thermal noise accurately.
Today the most prevalent method in use for gravitational wave detectors was developed by Yuri Levin~\cite{Levin:1998}, in which an oscillatory pressure resembling the spatial profile of the laser beam is imagined to be applied to the front surface of a mirror. The average power dissipated in the mirror due to this pressure then gives rise to mirror surface position fluctuations, again by applying the F-D theorem. 

There are different manifestations of thermal noise to be considered. In general, the term Brownian thermal noise describes noise arising from internal friction, which can come, for instance, from material defects. Internal friction within materials leads to damping of the harmonic oscillator modes, resulting in an increase of motion at the off-resonant frequencies of each mode. As a consequence, ultra-low-loss materials have to be used to minimise this motion, and the resonant frequencies (which still exhibit the highest motion) need to be shifted out of the observation band wherever possible.

Another class of thermal noise stems from thermoelastic dissipation, caused by temperature fluctuations or temperature gradients. When a suspension fiber bends, a temperature gradient at the bended flexure develops, which in turn leads to a dissipative heat flow that generates mechanical noise via the F-D theorem.
Statistical fluctuations of temperature, as caused by heat dissipation, for example, also cause mechanical displacement fluctuation via the linear thermal expansion coefficient of materials. Both of these processes are subsumed as thermoelastic noise.

Temperature fluctuations can also lead to refractive index fluctuations.
This effect is called thermorefractive noise. Thermoelastic and thermorefractive noises are particularly important for the optical coatings on the mirrors that provide the high reflectivity required for the test masses. They can be categorized together as thermo-optic noise, and they partially cancel each other out when treated coherently~\cite{Evans:2008}. 
Despite this partial cancellation, coating thermal noise is still the most relevant and limiting thermal noise source in current room-temperature detectors such as Advanced LIGO and Advanced Virgo. Research is ongoing to identify improved coating materials and designs, and also to identify coating options for other test mass materials in addition to cryogenic operation.

\subsubsection{Seismic and gravity gradient noise}
\index{seismic noise}\index{gravity gradient noise}

Seismic ground vibrations limit the performance of all terrestrial interferometric gravitational-wave detectors. They are a dominant noise source at frequencies below about $20$\,Hz and when they are particularly large, they can cause a loss of resonance of the Fabry-Perot cavities and significantly reduce the detectors' observing time. 
Seismic noise predominately originates from ocean and ground water dynamics, earthquakes, wind, and human-induced activities, as well as slow gravity drifts and the atmosphere. These ground and atmospheric disturbances couple to the test masses in two ways: by mechanically moving the mirror suspensions and by direct gravitational attraction.

\begin{figure}[ht]
\centering
\includegraphics[width=0.9\textwidth]{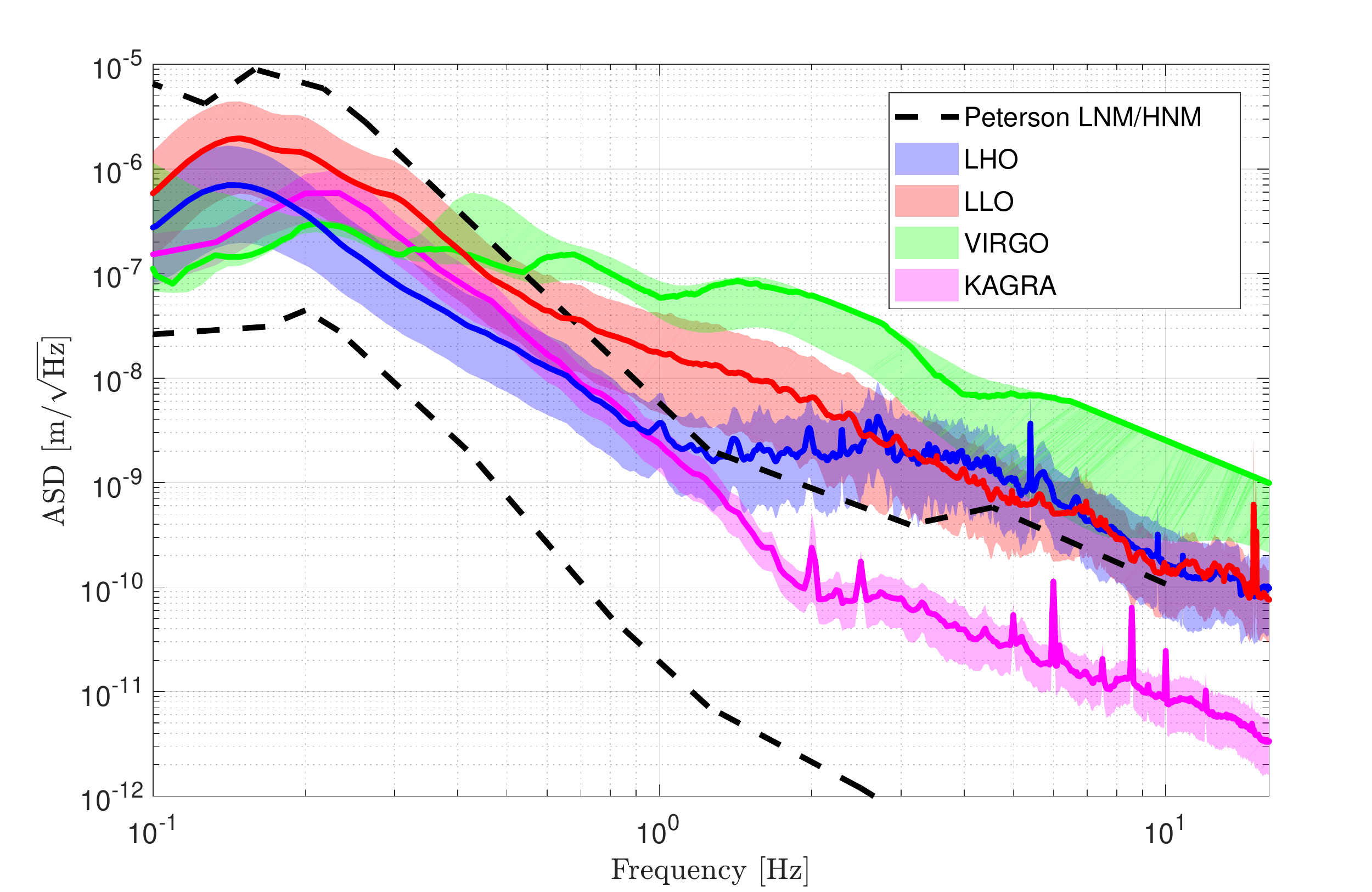}
\caption{Amplitude spectral density of seismic noise in the horizontal direction at each of the LIGO, VIRGO and KAGRA sites. The solid curves represent the mode, while the shaded bands show the $10^{\mbox{th}}$ to $90^{\mbox{th}}$ percentiles. The dashed curves represent Peterson's low and high noise models \cite{Peterson}.}
\label{fig:seis}
\end{figure}

Fig.\,\ref{fig:seis} shows the typical amplitude of ground displacements as a function of frequency at each of the advanced detector sites. The influence of these ground disturbances on the test masses is suppressed through the use of innovative vibration attenuation systems which decrease seismic noise by up to more than twelve orders of magnitude at frequencies above $10$\,Hz. These passive and active seismic isolation systems are described in Sec.\,\ref{sec:sei}.

The direct coupling to the test masses of density fluctuations in the atmosphere and in the ground is known as gravity gradient noise or Newtonian noise. Although gravity gradient noise cannot be directly attenuated, accurate models can be used to subtract its effect from the gravitational-wave detector output. 
Early analytical estimates of gravity gradient noise can be found in Weiss's and Saulson's work \cite{Weiss:1972,Saulson:1984}, and the most recent modeling is based on information from seismic surveys with large sensor arrays placed on the surface of the ground. Simulations on how to optimize the sensor array configuration to achieve a more accurate estimate of gravity gradient noise is an ongoing area of research.

Mitigation techniques for seismic noise such as the vibration isolation systems and gravity gradient noise subtraction need not be so heavily relied upon if the detector sites themselves could be carefully selected (which is important for potential future detectors as discussed in Sec.\,\ref{sec:outlook}) 
and if care is taken to minimize machine-induced vibrations, infrasound, or acoustics from turbo pumps, HVAC systems, water lines, etc. If one has the opportunity to visit a site, you may well be impressed by just how still it is inside the instrument halls that house the main interferometer components.
For a more in-depth review of environmental noise couplings to terrestrial laser interferometers, we refer the reader to the chapter on the subject within this volume.

\subsubsection{Noise from technical constraints}
Although the more fundamental noises (quantum noise, thermal noise and seismic noise) place ultimate constraints on detector sensitivity, 
a plethora of technical noises often limit what can actually be achieved in some frequency bands and a significant portion of interferometer commissioning is dedicated to reducing them. What we call technical noises, in loose distinction to fundamental noises, is everything that arises from the realities and imperfections of the technologies used. Imperfect sensors, the need for control systems, thermal effects, and cross-couplings of different kinds all create pathways for either displacement or sensing noises. Describing each specific noise source (as appears in the complete noise budget shown in Fig.\,\ref{fig:NB}) is not within the scope of this chapter, but we present here examples of some of the means by which technical noises come about:

\begin{itemize}
    \item \textbf{Feedback loops} (see Sec.\,\ref{sec:ctrl}) can impress the intrinsic noise of auxiliary sensors on the gravitational-wave readout channel. Even the best current efforts to suppress test mass motion locally with seismic pre-isolation systems leave the test masses with low-frequency motion around 1\,Hz or below, which is too large for interferometer operation. It therefore needs to be suppressed further by alignment feedback, using signals derived from the global operation of the interferometer. These feedback loops currently need a minimum bandwidth of a few Hz, which results in imprinting additional noise that stems form the sensing (shot) noise of these auxiliary sensors. Constraints on the stability of these feedback loops constrain the ability to filter this noise contribution from the gravitational-wave readout. 
    
    \vspace{2mm}
    \item \textbf{Cross-coupling of degrees of freedom} can be significant for several sub-systems because it provides a multitude of pathways for noise couplings. At the highest level, some of the technical noise sources that affect the strain sensitivity stem from sensing noise of auxiliary length degrees of freedom (length of the power- and signal recycling cavities, for instance) that cross-couple to the differential arm length. At a lower level, within some subsystems, the hardware itself can have mechanical cross-couplings. This shows up, for instance, in the multi-stage test mass suspension systems, where 
    cross-couplings between mechanical degrees of freedom cause some degrees of freedom, which may be less well isolated from ground motion, to add noise to others that are more relevant for the longitudinal test mass motion. Sensors themselves may also be imperfect in distinguishing degrees of freedom. Seismic acceleration sensors, for instance, are plagued by cross-coupling from tilt motion of the ground, which is currently a limitation to the performance of seismic pre-isolation systems. 
    Cross-coupling terms can be linear, bi-linear, or non-linear. A form of bi-linear coupling (involving two linear coupling parameters that multiply) is the residual motion of laser beam spot position on a test mass, which, when combined with test mass angular motion, creates a change in the longitudinal position of the test mass as sensed by the laser beam.

    \vspace{2mm}
    \item \textbf{Scattered light}\index{scattered light} can be generated by unevenness or dust contamination on the test mass surfaces. If this vagabonding light finds its way back into the main beam of the interferometer, it inserts a small additional phase shift, contaminating the gravitational-wave measurement in the form of sensing noise. Scattered light can also disturb the position of the test masses through the radiation pressure it exerts, which acts as a displacement noise. Substantial effort goes into the design and implementation of baffles to absorb scattered light before it does any harm and into further improving the optics to reduce the generation of scattered light. 
    
    \vspace{2mm}
    \item \textbf{High power} in the interferometer is technically very challenging to achieve because laser light absorbed by the optics creates thermal distortions, which in turn produce higher order spatial modes of the laser beam. These higher order modes can deteriorate the interferometer performance by creating new cross-coupling paths or contaminating sensing signals. 
    Thermal compensation systems have been developed to mitigate this problem and push the boundary of high power application. 

\end{itemize}

\subsection{Some enabling technologies}
\label{subsec:tech}

While noise sources set fundamental or technical limits on obtainable sensitivity for a given detector design,
enabling technologies are those that are required to construct an instrument of that sensitivity in the first place.
Here we briefly introduce some of these enabling technologies.

\subsubsection{Lasers}
The development of the laser\index{laser} in the 1960s
enabled decisive progress in interferometry. 
Because of their unique design, lasers produce very high intensity light, emitted as a bundled beam in one direction. In addition, laser light is monochromatic to a very high degree, which means it has a very well-defined wavelength, making it perfect for use in interferometers. Simply put, the more monochromatic the light, the easier it is to accurately measure its phase, which is the purpose of an interferometer. 

In the 1980s and early 90s, the most ubiquitous type of laser used for the research on gravitational-wave detectors was the argon ion gas laser. While output powers of up to $10$\,W were available, these lasers were not operationally reliable over longer periods of time, as was required for gravitational-wave observatories.
Substantial progress was made with the development of the nonplanar ring oscillator (NPRO) \cite{Kane:85}, a core element of laser-diode-pumped solid-state lasers. The NPRO\index{NPRO} design provides high reliability as well as very low intrinsic frequency and amplitude noise. For the widely used wavelength of 1064\,nm, neodymium-doped yttrium aluminum garnet (Nd:YAG) crystals are used as the lasing medium. 
Driven by the needs for higher power to reduce shot noise, as described in Section~\ref{sssec:quantum}, the NPRO based lasers have been extended
with single-pass amplifiers and/or injection lock cavity-based amplifiers to yield output powers of up to $200$\,W or more~\cite{RSG:2019}. 

While the intrinsic noise properties of the NPRO are excellent with respect to other lasers, the requirements for their use in gravitational-wave detectors demand active pre-stabilisation techniques.
Active feedback control is mandatory for stabilising the frequency and amplitude of the laser light, and an optical feedback-controlled resonator (pre-mode-cleaner) is used to improve the spacial purity of the laser mode.
Further stabilisation of frequency and amplitude noise is achieved by nested feedback control, using sensors within (or intrinsic to) the main interferometer \cite{Kwee:2012}. In particular, the km-long arms of a gravitational-wave detector with their seismically isolated mirrors provide the best possible frequency reference within the gravitational-wave frequency band of interest. The stabilised lasers thus achieve frequency noises down to micro-Hertz, 20 orders of magnitude below the lasing frequency.

Research is ongoing for the use of fiber-based laser amplifiers and configurations of ever higher output power,
as required for future gravitational-wave detectors.

\subsubsection{Vacuum systems}
\label{sec:vacuum}

Gravitational wave detectors feature comprehensive ultra-high vacuum (UHV)\index{UHV} systems which serve several purposes. 
First, fluctuations in the effective refractive index of residual gas 
can mask or imitate the expected gravitational wave signals. A high enough vacuum can reduce this phase noise from residual gas density fluctuations along the beam path to an acceptable level. Moreover, the vacuum environment isolates the test masses and other optical elements from acoustic noise, and reduces test mass motion excitation due to residual gas fluctuations. It reduces gas damping in the mirror suspensions, leading to lower suspension thermal noise, and contributes to the thermal isolation of the test masses and their support structures. Finally, the UHV environment contributes to preserving the cleanliness of optical elements.

\begin{figure}[ht]
\centering
\includegraphics[width=0.8\textwidth]{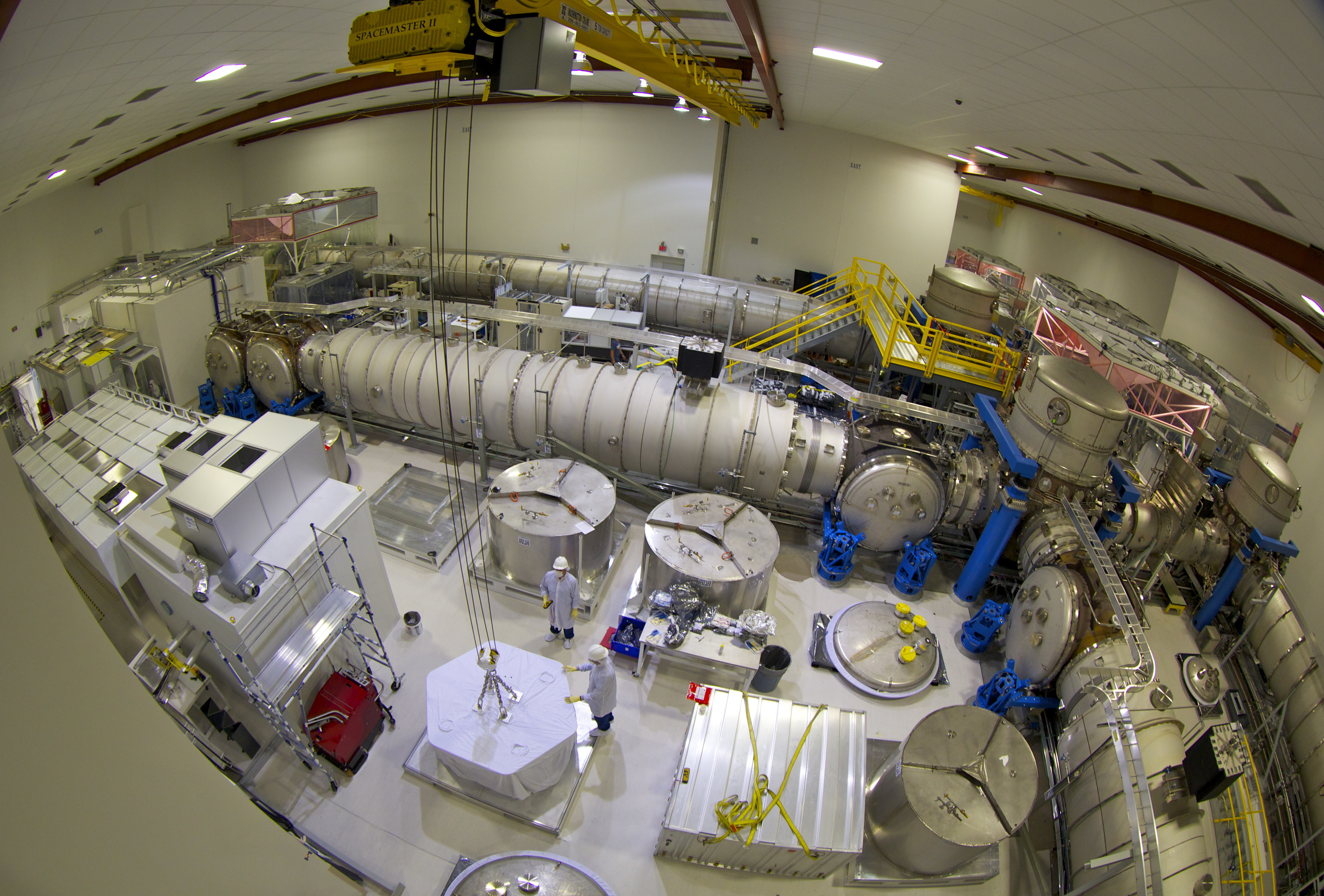}
\caption{The laser and vacuum equimpent area at the corner station of the LIGO Hanford Observatory. Photo courtesy: Caltech/MIT/LIGO Lab.}
\label{fig:LVEA}
\end{figure}

The vacuum systems of gravitational wave detectors are composed of two basic elements: chambers and beam tubes. The vacuum chambers house hardware including electrical, mechanical, and optical systems. 
The chambers are equipped with pumping stations and instrumentation for vacuum measurement and contain vacuum isolation valves and access doors as frequent access is required. The beam tubes connect the chambers over kilometer-long distances. Excellent vacuum must be maintained and the beam tubes are never vented. Figure \ref{fig:LVEA} shows the beam tubes of the LIGO Hanford detector emanating out from vacuum chambers that house the optics.

So-called cryolinks are used to connect the beam tubes to the corner stations. The cryolinks are big cryo-pumps for pumping water vapor and allow the interferometers to obtain pressures as low as several $10^{-10}$ mbar. Magnetically levitated turbo-molecular pumps are employed for initial evacuation only, while ion pumps assisted by non evaporable getters are used in normal operation. No rotating or vibrating machinery is permitted in the vicinity of the test masses during interferometer operation.

While the construction of the vacuum chambers is reasonably conventional, this is not the case for the beam tubes, which are highly customized. The impressive engineering achievement of constructing four 4\,km-long stainless steel tubes for LIGO---amongst the largest UHV systems in the world---required factory-like plants to be constructed at each of the Hanford and Livingston sites.
The LIGO vacuum tubes were produced by coil spiral-welding steel from rolls to tubes 1.2\,m in diameter and 16\,m long. Each 16\,m section was cleaned and leak-checked, and an FTIR analysis carried out to confirm that the tubes were free from hydro-carbons. The sections were then butt-welded together in the field using a traveling clean room, yielding over 50 linear km of weld. 
The raw stock material, stainless steel 304L of 3.2\,mm thickness, was air baked for 36 hours at a temperature of $455^\circ$C, resulting in a final hydrogen out-gassing rate of $< 10^{-13}$ Torr l/s/cm$^2$. 

The operation and maintenance of such a large vacuum installation over a period of decades constitutes a challenge. And due to its expense, the lifetime of the beam tubes largely determines the lifetime of a site.

\subsubsection{Seismic isolation}
\label{sec:sei}

The detection of gravitational waves by terrestrial laser interferometers requires the use of unparalleled vibration isolation systems\index{seismic isolation}. Given that typical displacements of the Earth's surface are of the order $10^{-9}$\,m$/ \sqrt{\rm Hz}$ at 10\,Hz (see Fig.\,\ref{fig:seis}), the main task of seismic isolation is to reduce mirror motion by more than 12 orders of magnitude. The effectiveness of the seismic isolation system ultimately sets the lower limit of the frequency band of terrestrial detectors. 

The high level of isolation of the mirrors from the ground
is achieved through the use of a series of passive and active control techniques. The mirrors are suspended as cascaded pendula, which, in turn, may be mounted on actively isolated platforms, in which a number of sensors on the ground and on the platform are used for feed--forward and feed--back control, respectively. The styles of seismic isolation in the various detectors vary by putting more or less emphasis on the extent of the passive versus active isolation.

\begin{figure}[ht]
\sidecaption
\includegraphics[width=0.6\textwidth]{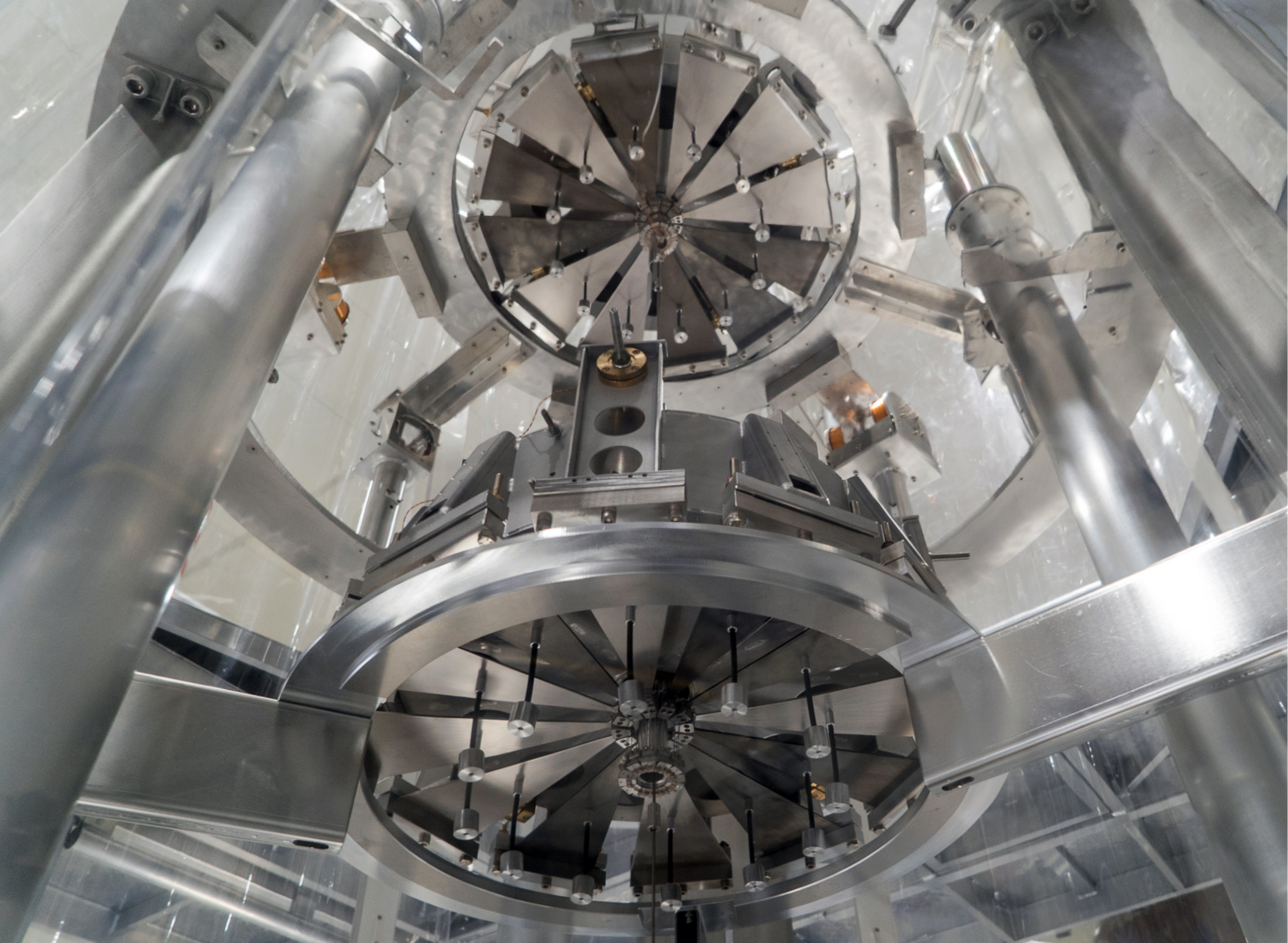}
\caption{Two of the five pendulum stages from which the mirror itself is suspended, as used in the Virgo detector. The full height of the multi-stage pendulum is an impressive 8\,m. Photo courtesy: Giacomo Raffaelli/Sonic Somatic}
\label{fig:sa}
\end{figure}

The most critical components of the seismic isolation system are the pendulums\index{pendulum}. Suspension of the mirrors provides attenuation from the ground above the pendulum's resonance frequency that scales as $f^{2N}$, where $N$ is the number of pendulum stages. Making the pendula long lowers their resonance frequency, while more stages produce a stronger reduction of seismic noise towards higher frequencies. This suffices for bringing the relative motion of the mirrors at frequencies above the pendulum resonance down to levels where seismic noise no longer dominates; however, the unsuppressed and even amplified ground motion at frequencies at and below the pendulum resonance remains a challenge. 
In order for the interferometer to stay within its linear operating range, the relative root-mean-square motion over all frequencies of the suspended mirrors must be reduced by over six orders of magnitude. 
This is to a large part achieved through global feedback control systems that act on the mirror suspensions, but also through the use of seismic pre-isolation systems, from which the pendulums themselves are suspended. The latter may take the form of an inverted pendulum as in Virgo, or an actively controlled platform on springs as is used in LIGO. Details about the design, operation and performance of the seismic isolation systems can be found in \cite{Matichard:2015,Accadia:2011}. 

Finally, because pendula chains have six degrees of freedom per stage, there is ample motion at the eigenmode frequencies of the suspension if left uncontrolled. Systems that are local to each suspension chain are thus also required to actively damp the suspension eigenmodes with feedback control.

\subsubsection{Feedback control}
\label{sec:ctrl}

The use of feedback control\index{feedback control} is an essential aspect of building laser interferometers capable of detecting gravitational waves. 
We have already mentioned the stabilisation of the lasers and local control of the suspensions above; here, we describe more broadly how feedback is implemented in the detectors.

To facilitate the measurement of the tiny displacements that the main interferometer can measure, 
feedback control is required to hold all optical resonators and the Michelson interferometer at their operating points to within a small fraction of the laser wavelength.
This enables both the build-up of the required light power in the interferometer arms and ensures a linear response to residual deviations of arm length, which contain the gravitational-wave signal. 
To achieve this, the length sensing and control system must control five length degrees of freedom and keep all mode cleaner cavities on resonance. 

In addition to these length degrees of freedom which are related to the distances between the optics and
the laser frequency, all suspended optics must also be controlled in two angular degrees of freedom, pitch and yaw, which correspond to the vertical and horizontal directions of a reflected beam.
All of these length and alignment degrees of freedom come on top of the local controls of the suspensions and constitute the global control of the interferometer. Although the interferometer is an analog instrument, it is interfaced through a digital control system which allows complex filters to be implemented and tuned from the control room in order to create control signals.

The various length and angular degrees of freedom are sensed primarily through the use of radio-frequency sidebands\index{sidebands} on the carrier light which are created through phase modulation by electro-optic modulators before entering the main interferometer. The differential arm length signal, which is sensitive to gravitational waves, is currently sensed using homodyne readout at the anti-symmetric port\footnote{All detectors initially used heterodyne readout, where the gravitational wave sidebands beat against radio-frequency sidebands generated at the input of the interferometer.}, where the gravitational wave signal sidebands are converted to power variations of the light. To generate suitable error signals, extensions of the Pound--Drever--Hall method of laser frequency stabilization are used \cite{Drever:1983, Black:2001, Strain:2003}. 

While the interferometer is locked, a steady state in which all cavities are close to their nominal operating points, feedback control loops are based on the assumption of linear time independent (LTI) systems. This is not the case, however, during the lock acquisition phase when steady state control signals for all optical degrees of freedom do not yet exist.
Feedback control for lock acquisition has always been a particular challenge, 
with dedicated methods having to be developed. 
One example is the strategy of using auxiliary laser systems to first lock the Fabry-Perot arm cavities before handing the lock off to the main laser.

\subsection{Things not yet mentioned}
We have insufficient space here to comprehensively list all
critical technologies or instrument sub-fields, but mention briefly a few more here.

\subsubsection{Precision optics}
Terrestrial laser interferometers require mirrors that perform at the edge
of current technology in terms of reflectivity, scattering of light, and losses, and in several cases have pushed the limits of what can be achieved. The main requirements for the primary optics, the test masses, are the following: precisely polished surfaces with root-mean-square roughness of less than 1\,nm over the full face of the $\sim30$\,cm diameter optic; absorption at or below the 1\,ppm range; and optical coatings with reflectivities as high as 99.995\,\% or more.

Such surface precision is necessary in order to minimise the amount of light that gets scattered into spatial modes that are lost to the interferometer, or worse, contribute to excess noise. Low absorption is needed to avoid or minimise thermal distortion of the mirrors, which in turn would generate higher order spatial modes. High reflectivities are necessary to facilitate sufficient resonant enhancement of laser light to achieve a high total power in the interferometer arms.

\begin{figure}[ht]
\sidecaption
\includegraphics[width=0.6\textwidth]{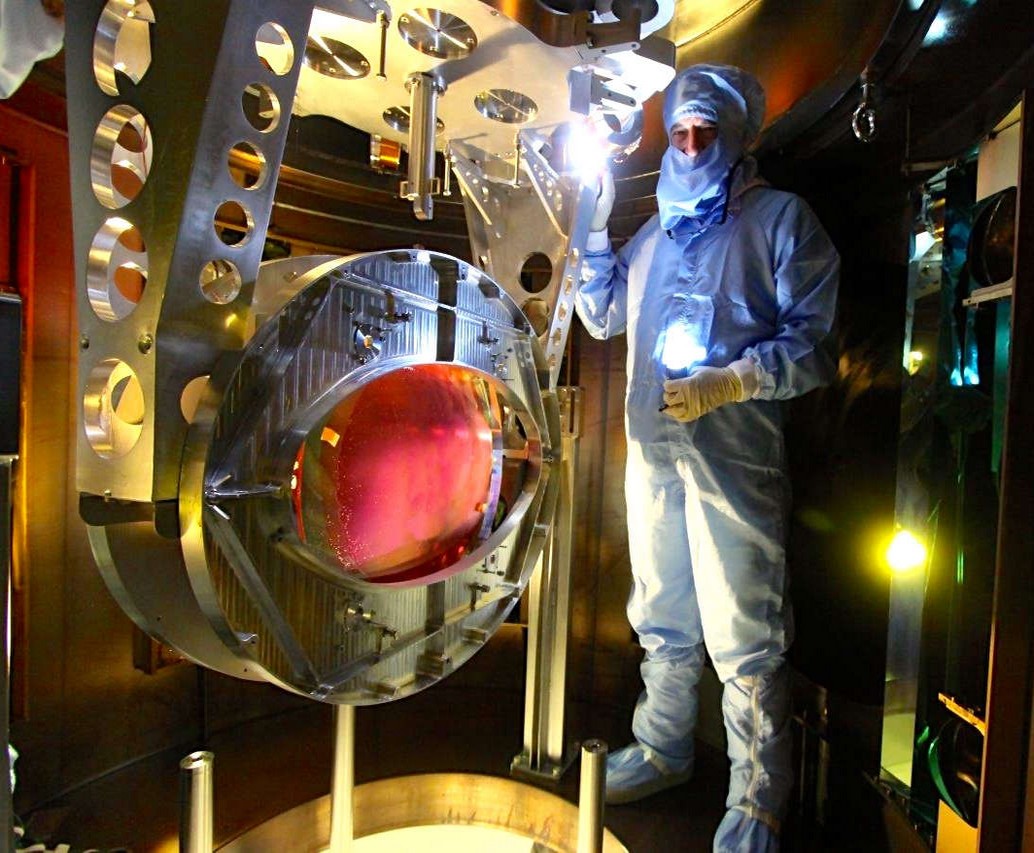}
\caption{The beam splitter used in the Advanced Virgo gravitational wave detector. The beam splitter is framed by a metal screen,
which absorbs scattered light\index{scattered light}. 
Photo courtesy: EGO/Virgo.}
\label{fig:vbs}
\end{figure}

One of the most critical parts of the precision optics are the optical coatings, which are put onto the mirror substrates using an ion-beam-sputtering process. 
The design of the coatings determines not only the mirror reflectivities,
but also the thermal noise properties of the coatings, which is a dominant noise sources as pointed out in Sec\,\ref{sssec:thermal}.
The polishing process and coating uniformity both determine the scattering
of light. Excess scatter and absorption can occur from so-called point defects
or dust particle contamination, such that extreme cleanliness procedures
on fabrication, handling and installation of the optics are mandatory.
Figure~\ref{fig:vbs} shows the beam splitter of the Virgo interferometer
during installation.

\subsubsection{Simulation and diagnostic methods}
A wide range of simulation tools have been developed
by the gravitational-wave instrument community, which are indispensable for the design and commissioning of the complex instruments.

A simple classification scheme splits the codes into time-domain and
frequency-domain tools. While time-domain tools can be the
most complete in simulating almost any physical system property,
this comes at the cost of computational load. On the contrary, frequency domain tools are much faster, but are limited to steady-state applications and linear-time-invariant (LTI) systems. Widely used frequency-domain simulation packages for simulating optical systems in gravitational wave detectors include Finesse, Optickle, and MIST\,\cite{Finesse:2015,Optickle:2015,Vajente:2013}.
A third class of code uses propagation of Gaussian optical beams
with FFT-based methods,
which is useful to estimate the effect of mirror imperfections.
While all of the above are numerical models, analytical modeling tools 
have also been developed to study basic noise properties critical for interferometer design.

Diagnostic methods to investigate properties of the running laser interferometer and its sub-systems are likewise diverse. A major aspect of the interferometer commissioning work is the identification of noise sources that couple to the main gravitational-wave readout.
A workhorse technique to estimate these noise couplings is to inject enough artificial noise into some sub-system such that this
noise is dominant in the gravitational wave readout. This allows the measurement of the transfer function
(the frequency dependent coupling coefficients) from the sub-system to the gravitational wave readout.
In a second step, the noise spectrum of the sub-system is measured without the injected noise and then multiplied by the transfer function. The result is a \emph{noise projection} of the sub-system to the gravitational wave readout. 

\begin{figure}[htb]
\includegraphics[width=\textwidth]{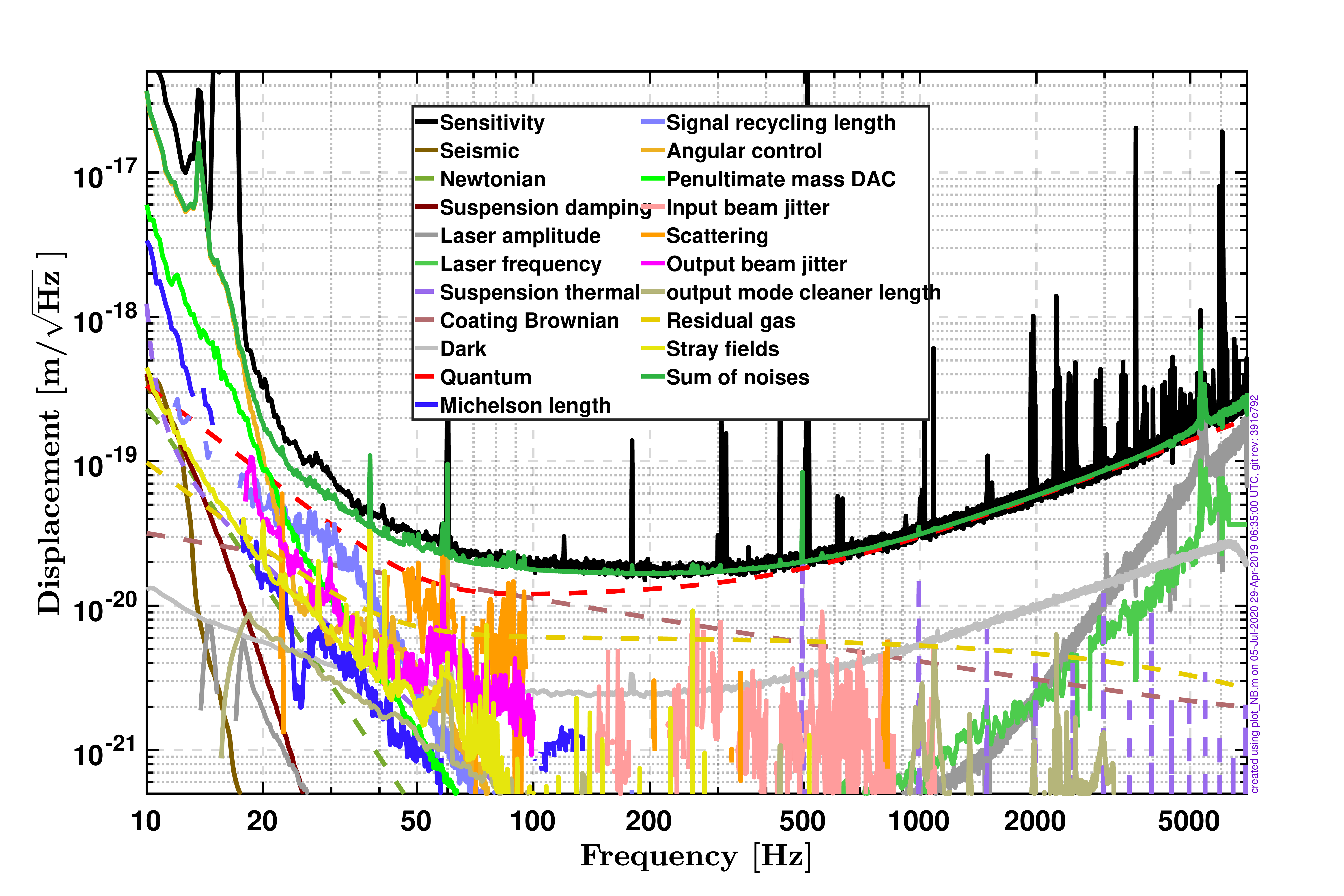}
\caption{Noise budget of the Advanced LIGO gravitational wave detector \cite{LIGOO3Sensitivity}. Solid lines show measured noise projections and dashed lines are models. The incoherent sum of all known noise sources matches the measured noise curve at nearly all frequencies.}
\label{fig:NB}   
\end{figure}

Fig.\,\ref{fig:NB} shows a recent noise budget\index{noise budget} for the Advanced LIGO detector, which includes measured noise projections of over a dozen sub-systems and several modeled noise sources as well. Figures such as this inform further commissioning or noise mitigation steps. While this noise projection method typically assumes steady-state LTI systems, in some instances non-linear or bi-linear couplings need to be taken into account and time-domain analysis is required to study transient excess noise events in the instrument.

\subsubsection{Robustness}
The concept of robustness addresses the general question of how resilient the operation of the interferometer and its performance are against intrinsic or extrinsic conditions or disturbances. A typical example is the question of how robust the interferometer operation is against elevated seismic noise or remote earthquakes. While robustness may be increased by stronger actuators to counter external motion, sensitivity may be degraded by such measures.
An which stems from high laser power in the interferometer, is the so-called parametric instabilities that can inhibit operation. Parametric instabilities are excitations of mechanical eigenmodes of the test masses in a run-away positive feedback loop driven by the high-power laser field impinging the mirrors \cite{PI:2019}. Active or passive damping of the mirror modes has to be employed to keep this problem at bay.
Finally, procedures have to be designed and commissioned for lock acquisition, the process of bringing the multiple optical resonators reliably to their operating points. Lock acquisition is a complicated problem because global control signals for all degrees of freedom do not exist before cavities are close to their operating points. Methods and sub-systems are devised with the sole purpose of bringing interferometers close to their operating points in an automated sequence of steps.

\subsubsection{Calibration}
Calibration of the output signal of a laser interferometer in terms
of gravitational wave strain is obviously important for astrophysical
inferences on the observed data. Calibration is typically performed by applying a known displacement to one or more test masses and inferring strain sensitivity from the known arm length of the detector. Since the detector's optical response function to displacement and strain variations is typically frequency dependent, calibration has to be performed at multiple frequencies. Techniques to apply well estimated displacements to test masses comprise voice-coil actuation, radiation pressure actuation with auxiliary lasers, or even gravity actuation. A challenge for the future is to achieve sub-percent accuracy on calibration, as required for some astrophysical inferences \cite{Sun:2020,Maggiore:2020}.

\section{Laser Interferometers World-Wide}
\label{sec:detectors}
Following the development of precision laser interferometry techniques in the 1970s and 1980s, several groups around the world began to develop detailed plans and to secure funding for kilometer-scale gravitational-wave detectors, ones large enough to have the promise of unambiguously detecting gravitational waves one day. Collaborations formed within the US to build and operate three 4-km long laser interferometers, known as the Laser Interferometer Gravitational-Wave Observatory (LIGO); Italy and France combined efforts to build a 3-km long detector called Virgo near Pisa, Italy; groups in Germany and the UK laid the groundwork for what would become GEO\,600, located in Germany; and the Japanese followed suit, constructing ever larger and larger prototypes until funding was secured in 2010 for a 3-km long underground detector, KAGRA. 

Success finally came in 2015, when Advanced LIGO, the current incarnation of LIGO that followed from a five-year-long program of instrument upgrades, captured the first-ever gravitational wave signal, produced by coalescing black holes \cite{GW150914}. As soon as the similarly-upgraded Advanced Virgo became operational, it too immediately began to contribute to the ever-increasing list of gravitational wave detections, most notably that of a binary neutron star inspiral, which was subsequently also observed by electromagnetic telescopes, ushering in the era of multi-messenger astronomy \cite{GW170817}.

In the following sections we highlight some key developments in the histories of these major gravitational wave detectors around the world and some of the main features that are unique to each instrument. Table\,\ref{tab:overview} provides a snapshot of the most critical design parameters of each detector.
We note that the history documenting the beginnings of the field of gravitational wave detection in the U.S. is a topic that is well-covered in a number of sources, for example \cite{Saulson:2019}, whereas the stories of the development of the detectors in Europe and Japan are perhaps a bit less well known.

\begin{table*}
\centering
\caption{Select design properties of each of the four gravitational wave detectors. DRFPMI stands for dual-recycled Fabry-Perot Michelson.}
\begin{tabular}{l l l l l}
\hline
           & Advanced LIGO & Advanced Virgo & GEO\,600 & KAGRA \\
\hline
arm length & 4 km & 3 km & 2$\times$600 m & 3 km \\
power recycling gain & 44 & 39 & 900 & 11 \\
arm power  & 800~kW & 700~kW & 20~kW & 400~kW \\
\# of pendulum stages & 4 & 8 & 3 & 6 \\
mirror mass & 40~kg &42~kg & 6~kg & 23~kg \\
mirror material & fused silica & fused silica & fused silica & sapphire \\
temperature & room & room & room & cryogenic \\
topology & DRFPMI & DRFPMI & DRMI & DRFPMI \\
location & surface & surface & surface & underground \\
\hline
\end{tabular}
\label{tab:overview}
\end{table*}

\subsection{LIGO}\index{LIGO}

The beginnings of the LIGO project date back to the early 1980s and were marked by the collaboration of Rainer Weiss\index{Weiss, Rainer} and Kip Thorne\index{Thorne, Kip} and a 1983 study known as the ``Blue Book" \cite{BlueBook} in which Weiss and his collaborators make the case for the feasibility and timeliness of building long baseline laser interferometers, writing:
\begin{quotation}
The positive conclusion of this study may have been anticipated. It could have been otherwise: The basic concept could have been flawed, the technology could have been inadequate, the costs could have been beyond reasons. None of these appears to be the case.
\end{quotation}

The National Science Foundation (NSF) provided funding for development work and Weiss, Thorne and Drever jointly managed the new LIGO project. However, due to their differing viewpoints on technical issues and styles of management, their collaboration lasted only three years. 
Progress was accelerated when the NSF\index{National Science Foundation} urged the appointment in 1987 of a single director to the LIGO project. One of the important contributions of the new director, Rochus Vogt, which helped to propel the project forward was his executive decision that the interferometers should use Fabry-Perot resonators\index{Fabry-Perot resonator} to increase the effective path length of the arms as proposed by Drever, rather than the delay line\index{delay line} technique favored by Weiss.

By 1989, LIGO scientists had submitted an application for funding to the NSF, proposing the construction of two geographically distant facilities in the US. One site would house a single 4-km long interferometer and the other would house two interferometers: one 4\,km and the other 2\,km. The two co-located interferometers would serve as a means to rule out local disturbances as the cause of candidate gravitational wave signals. Due to the uncertainty of the maturity of the technology required to construct detectors of this scale, however, the first interferometers to be installed would remain technically conservative. The proposal thus included a plan to later install a more technically advanced interferometer that would have better chances of detecting gravitational waves.

The discussions about whether to finance Initial LIGO---as it would later be called---lasted for several years. There was considerable opposition, particularly from astronomers. They were irritated by the fact that LIGO described itself as an observatory despite the small chances of detecting any gravitational waves, and they feared they would lose the money allocated for their own projects. The project was approved in 1990 and over the next years the US Congress appropriated funds specifically for LIGO construction, such that the NSF would not have to withdraw too much support from other projects. The NSF selected Hanford, Washington and Livingston, Louisiana, as the sites for the LIGO observatories. 

Barry Barish and project manager Gary Sanders took over the management of the LIGO project in 1994, the same year that construction of LIGO's buildings and vacuum systems began. Both Barish and Sanders had extensive experience with large projects in particle physics and understood the necessity of strict management to successfully operate large-scale facilities. Under their lead in 1997, the LIGO Scientific Collaboration (LSC)\index{LSC (LIGO Scientific Collaboration)} was founded to organize the expertise amongst different research groups and to foster the national and international collaboration necessary for the successful development of the required technologies.

\begin{figure}[ht]
\includegraphics[width=\textwidth]{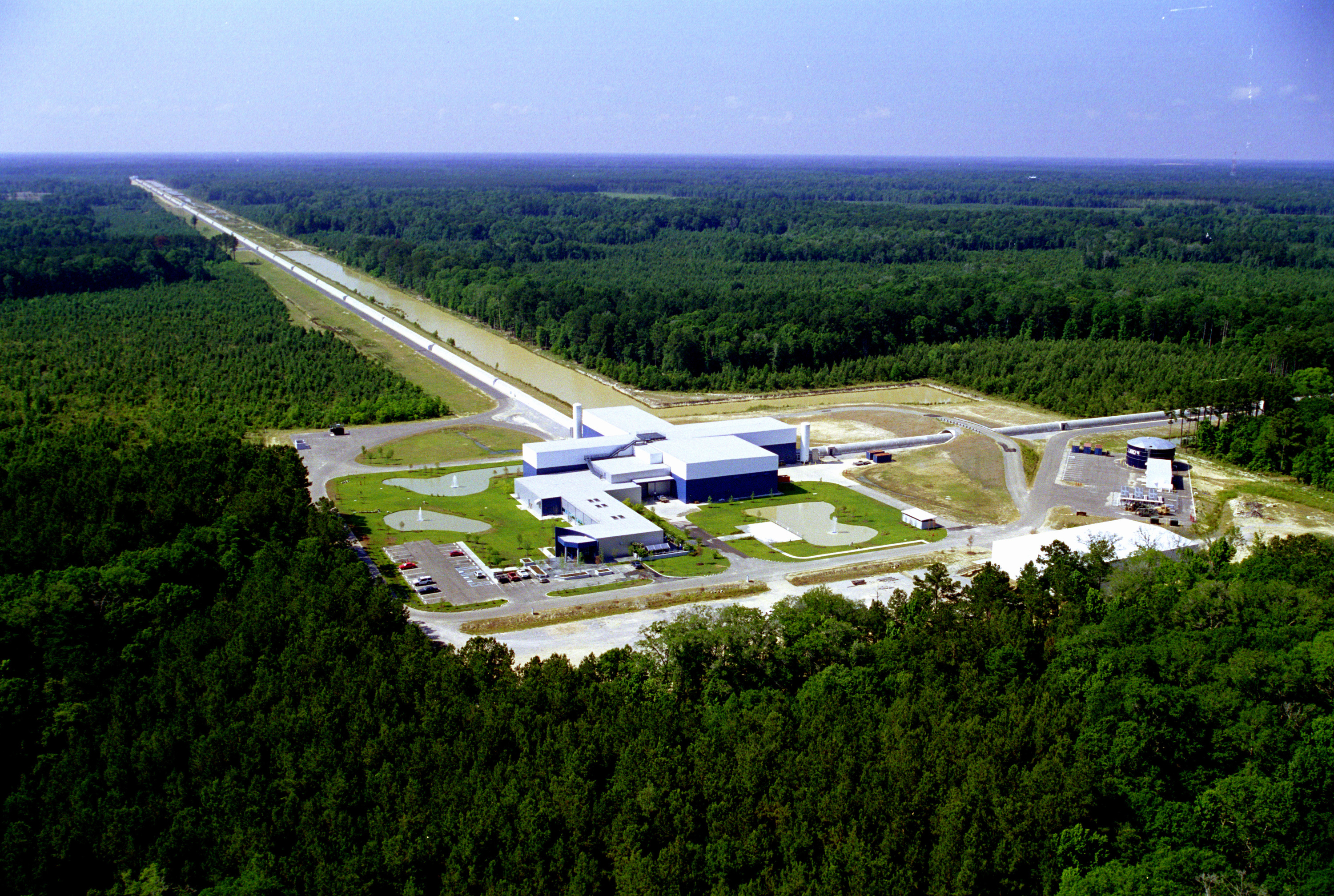}
\caption{The LIGO Livingston Observatory in Livingston, Louisiana, USA. Photo courtesy: LIGO Laboratory.}
\label{fig:LLO}
\end{figure}

The impressive engineering achievement of constructing LIGO's vacuum system was completed in 1998. In the several years that followed, power-recycled\index{power recycling} Michelson interferometers\index{Michelson interferometer} with Fabry-Perot resonators\index{Fabry-Perot resonator} in each arm would be installed. Livingston housed one 4\,km long interferometer and Hanford housed two: one 4\,km in length and the other 2\,km long. In 2000 the 2-km interferometer at Hanford was the first of the LIGO detectors to be locked. Figure~\ref{fig:LLO} shows an aerial view of the LIGO site in Louisiana in 2015.

Intense years of commissioning the instruments followed in order to integrate all of the sub-systems, develop automated locking sequences, and most importantly, track down noise and reduce it in order to reach the design sensitivity. Commissioning was interspersed with a series of Science Runs in which the three Initial LIGO detectors collected scientific data. This culminated in the 2-year-long S5 run from 2005 to 2007. Although no gravitational waves were detected, upper limits on potential sources were placed, as with the Crab pulsar for example \cite{CrabPulsar}, and importantly, the maturity and understanding of the instrumentation proven solid.

Research and development work on Advanced LIGO\index{Advanced LIGO}, the interferometer that was to replace Initial LIGO, began around 1999 and involved scientists from the British and German GEO collaboration and from Australia. The main features of Advanced LIGO included the following upgrades: a more powerful 200\,W laser, compared to the 10\,W of Initial LIGO; larger and heavier test masses weighing 40\,kg rather than 10\,kg; welded suspension fibers made of glass rather than stainless steel; the replacement of single stage suspensions for the test masses with quadruple pendulum\index{pendulum} suspensions; active seismic pre-isolation of the test mass suspensions; electrostatic actuators; the addition of signal recycling\index{signal recycling}; and a change from heterodyne to homodyne readout. An intermediary project, Enhanced LIGO, implemented a select subset of these upgrades, which improved the sensitivity of the instruments by 30\% over Initial LIGO and the final Science Run, S6, was carried out from 2009 to 2010.

Financing of \$205M for three 4\,km long Advanced LIGO interferometers was secured in 2008 from the NSF, as well as in-kind contributions worth \$15M from the GEO collaboration. The original intent was to install one interferometer in Livingston and two in Hanford. But this would later be modified following additional studies of the benefits of having widely separated detectors for improving source localization (see Ch.\,5) compared to the better stochastic background measurements afforded by co-located detectors. The second Hanford interferometer is thus slated today to be installed in a new facility currently under construction in India and is called LIGO-India\index{LIGO-India} (see Sec.\,\ref{ssec:LIGO-India}). 

The construction of the Advanced LIGO interferometers took about four years and was completed in 2014, which then marked the start of the commissioning\index{commissioning} phase. Rapid progress could be made in improving the sensitivity at frequencies above around 100\,Hz, largely as a result of the experience from the first generation of detectors which had greatly helped inform the design of Advanced LIGO and the level of testing of subsystems before installation. Progress has been slower, however, below 100\,Hz, a region that is now possible to explore given the much-improved seismic isolation system. Challenges include electrical charges on the test masses that interact with electromagnetic fields in the environment and scattered light\index{scattered light}, amongst others. We refer the reader to the corresponding chapter in this Handbook on the topic.

The objective for Advanced LIGO\index{Advanced LIGO} was to improve strain sensitivity by a factor of 10 compared to Initial LIGO\index{Initial LIGO}. About one third of this improvement was achieved by both Advanced LIGO detectors by the beginning of their first observation run (O1) in September 2015. This corresponded to a range of about 60\,Mpc for the detection of binary neutron star mergers\index{binary system} and proved sufficient for the first gravitational-wave detections. As a result of continued commissioning during breaks between observing runs, the Advanced LIGO detectors reached a sensitivity of 120\,Mpc (LHO) and 140\,Mpc (LLO) for the O3 run of 2019--2020.

\subsection{Virgo}\index{Virgo}

The idea of creating Virgo traces back to a discussion between Alain Brillet and Adalberto Giazotto during the 4\textsuperscript{th} Marcel Grossman Meeting in 1985 during a walk in the courtyards of the University Sapienza in Rome. Both had been working on aspects of gravitational wave detection for some time already.

Alain Brillet is a French expert in optics
who became interested in laser interferometric gravitational wave detectors during his stay at the University of Colorado in Boulder (1977--1978), where he came in contact with Peter Bender, who, together with Jim Faller, would later propose the basic concept behind LISA \cite{bender_faller}. After discussions with Ron Drever, Albrecht Rüdiger, Roland Schilling and Rai Weiss, he started a program at Orsay in collaboration with Philippe Tourrenc and Jean-Yves Vinet. The French team, which included Nary Man and David Shoemaker, focused in the early 1980s on how to reduce the dominant noise source at frequencies around $1$\,kHz: shot noise. This required an enhancement of laser power stability and identification of an appropriate laser source. Brillet's team showed for the first time the efficiency of power recycling, a technique that was invented by Ron Drever and Roland Schilling, by demonstrating that the sensitivity of a Fabry-Perot Michelson interferometer is shot-noise-limited \cite{Man}. Moreover, they studied the use of Nd:YAG lasers as a replacement for the Argon lasers that were in common use at the time.

Adalberto Giazotto was an Italian particle physicist 
working at CERN and Daresbury. He became interested in gravitational wave research after the discovery of many new pulsars by radio telescopes. In the early 1980s the general consensus was that supernova explosions would be the prime sources of gravitational waves, and that gravitational wave detectors, such as the bar detectors of the time, should target frequencies around 1\,kHz.
Giazotto brought about a paradigm shift in the nascent field by focusing his attention onto pulsars as gravitational wave sources, with frequencies as low as $10$\,Hz. Detection of gravitational waves at such low frequencies presents formidable experimental challenges due primarily to seismic noise. It required designing and building a new type of vibration isolation system capable of attenuating seismic noise by more than twelve orders of magnitude starting at around $10$\,Hz. 

In the early 1980s Giazotto’s team started an experimental activity called IRAS (Interferometro per la Riduzione Attiva del Sisma) in San Piero a Grado in Pisa, Italy to develop a single degree of freedom seismic attenuation system based on a $1$\,m pendulum. This was followed by the development of the so-called super-attenuator, 
the multi-pendular vibration isolation system in use at Virgo today, which is capable of vibration isolation in all six degrees of freedom \cite{DelFabbro:1987}. To facilitate attenuation of seismic vibrations at as low a frequency as possible, the system uses anti-spring technology. Inverted pendula are employed as a first stage to suppress vibrations in the horizontal direction, and magnetic anti-springs counter vertical vibrations. Each mirror is suspended by wires from the last pendular stage.

The first proposal by the Italian-French collaboration 
for a km-scale gravitational wave detector, titled \emph{``Antenna interferometrica a grande base per la ricerca di Onde Gravitazionali,"} was prepared by the Pisa and Orsay groups in 1987. The funding request was written in Italian and submitted to INFN (National Institute for Nuclear Physics), the coordinating institution for nuclear, particle, theoretical and astroparticle physics in Italy. However, because there weren't any 
French particle physics groups involved, the French state research organisation CNRS (French National Centre for Scientific Research) could not support the proposal. 

After the initial collaboration between Orsay and Pisa, the project attracted several French and Italian groups, and in 1989 a detailed proposal with the title \emph{``The Virgo Project"} was submitted to both CNRS and INFN. This proposal already contained the agreement between group leaders from Italy, France, Germany, Scotland, and the USA to exchange all information and to collaborate on all aspects of the construction. Moreover it was decided to use the same standards for the data acquisition and data format for Virgo and LIGO (and future gravitational wave observatories not yet envisioned).
In 1992 the Virgo Project was officially approved by the French Minister Hubert Curien. A year later INFN 
gave its approval and the final agreement between CNRS and INFN was signed in 1994.

\begin{figure}[ht]
\includegraphics[scale=.27]{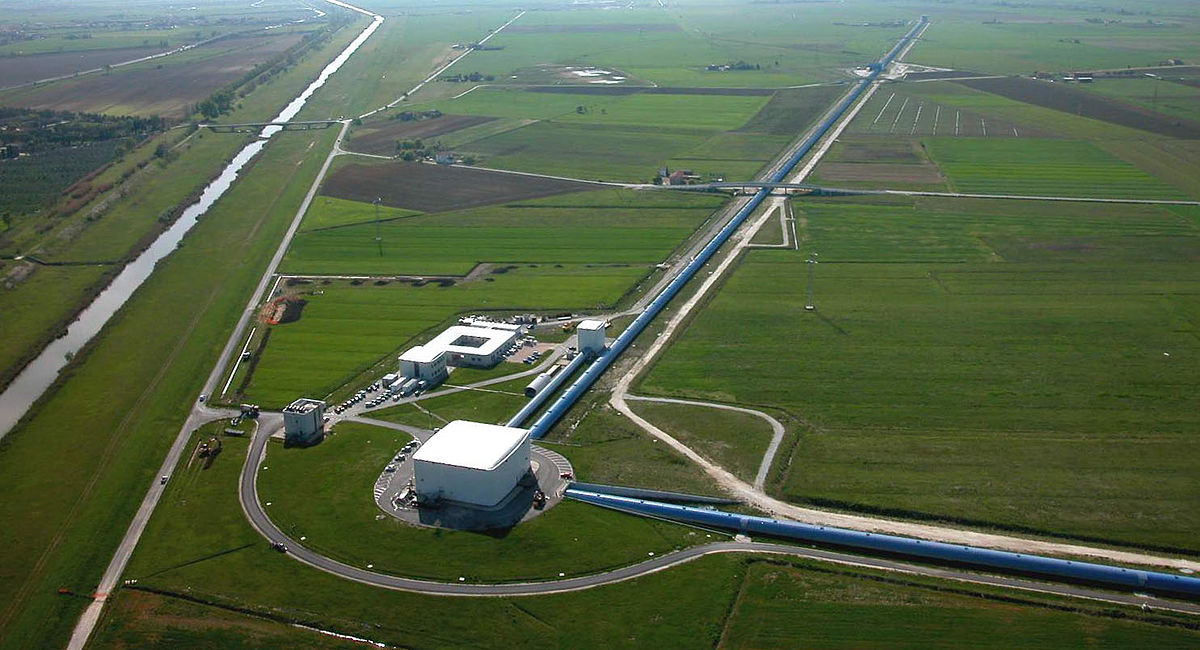}
\caption{The Virgo gravitational wave detector is hosted at EGO near Pisa, Italy. Each interferometer arm is 3\,km long. Photo courtesy: EGO/Virgo.}
\label{fig:virgo}
\end{figure}

Finding a site for Virgo was one of the most important issues. Giazotto took the initiative and proposed in 1988 the area of Tombola, inside a protected estate of pine groves located between Pisa and the sea. This located was ruled out, and instead the mayor of the town of Cascina, near Pisa, proposed in 1989 to realize the project in their territory. This land, about 350,000 m$^2$, was not public, however, and it took until 1996 to expropriate. After land acquisition was completed, construction of the Virgo facilities started in 1997 with Brillet and Giazotto as (alternating) Project Leaders. Fig.\,\ref{fig:virgo} shows a photograph of the Virgo site.

The Virgo interferometer was named after the Virgo Cluster, a collection of about $1,500$ galaxies in the Virgo constellation, about 50 million light-years from Earth. The construction of the Initial Virgo detector was completed in June 2003 and this was followed by commissioning. 
Since 2007, Virgo and LIGO Scientific Collaborations have agreed to share and jointly analyze the data recorded by their detectors and to jointly publish their results (an agreement to include KAGRA was signed in October 2019). Four science runs took place between 2007 and 2011, partly in coincidence with LIGO. In 2010 the original suspension steel wires were replaced by glass fibers in order to reduce the thermal noise. This was followed by a science run where Virgo was the first interferometer to use the monolithic suspension technology, originally developed for GEO\,600, with 40\,kg mirrors. 

In September 2011, the operation of Virgo was halted to begin the installation of the Advanced Virgo \cite{AdV} upgrade project that aims, like Advanced LIGO, for a factor of 10 improvement in sensitivity. Larger beams and heavier test masses of higher quality were employed. Highly uniform coatings of novel Bragg-reflector materials were realized by the Laboratoire des Matériaux Avancés (LMA) in France to decrease the optical losses and to increase the sensitivity of both Virgo and LIGO. The finesse of the main arm cavities was increased to allow the storage of up to 700\,kW of laser power, leading to a reduction of shot noise and better sensitivity at high frequency. Signals to control the various cavity degrees of freedom were obtained by direct digital demodulation, and a calibration method based on local variations of the Newtonian gravitational field was developed.
In 2019, the use of squeezed light in Advanced Virgo was realized together with scientists from the Albert Einstein Institute in Hannover, Germany. 

Following the installation of new hardware, the commissioning process for Advanced Virgo started in 2016, and in May and June 2017 Advanced Virgo joined the two Advanced LIGO detectors for a first ``engineering" observing period. On 14 August 2017, LIGO and Virgo detected a signal, GW170814, which was the first triple-detection of a binary black hole merger \cite{GW170814}. The addition of Virgo to the global network allowed for more sky coverage due to the complementary antenna patterns. For the first time determination of gravitational wave polarization became possible, and the distance measurement was improved. Also, the three-fold coincidence measurement resulted in increased robustness of gravitational wave detection, additional coincidence data, and provided an improvement in the ability to infer the properties of the gravitational wave sources. The significantly better sky localization of sources enabled the first detection of the merger of a binary neutron star, GW170817 \cite{GW170817}.

Advanced Virgo, together with Advanced LIGO, started their third observation run ‘O3’ on April 1, 2019 with an interruption in October 2019 to allow for maintenance and commissioning of the detectors. Although O3 was scheduled to take one year of data, it was suspended due to the COVID-19 pandemic on March 27, 2020 to guarantee the safety of personnel at the observatories. 
No less than 57 candidate events were collected, showing the scientific relevance of Virgo in the global network of second generation gravitational wave detectors. 

The operation and maintenance of Virgo is managed by a French-Italian consortium called EGO (European Gravitational Observatory), which was founded in 2000 and is governed by Italian law. Besides overseeing Virgo, EGO also serves a role of more generally promoting experimental and theoretical research of gravitational waves in Europe. EGO promotes relations between scientists and engineers, the dissemination of information and the advanced training of young researchers. Nikhef in the Netherlands joined the EGO Consortium in 2009 as associate member.
The Virgo Collaboration, whose members are responsible for designing, building and commissioning the detector and analysing the data, currently consists of about 600 members from 110 institutes in 12 European countries.

\subsection{KAGRA}\index{KAGRA}

Japanese scientists have been working on gravitational wave detection for decades. In the early 1990s the 20 prototype interferometer was built at the Mitaka campus of the National Astronomical Observatory of Japan as a project of the gravitational wave research group in Japan. This was followed by TAMA\,300, a gravitational wave detector with 300-m long interferometer arms that began operation in Mitaka, a suburb of Tokyo, in 1998. At that time TAMA\,300 was the largest and most sensitive detector in the world \cite{TAMA:2001}. The optical configuration used Fabry-Perot resonators in the arms, as well as power recycling. In 2000, a new world record for strain sensitivity of $10^{-21}$ was set, with TAMA\,300 being more than ten thousand times as sensitive than Weber's cylinder. In 2003, a joint data run with LIGO was accomplished.  However, the interferometer arms were too short to make any detections, and the anthropogenic seismic noise in Tokyo affected operations.

In 1999 the 20 prototype interferometer was used in the LISM (Laser Interferometer Gravitational-wave Small Observatory in a Mine) project and moved 1000\,m underground to Kamioka in Hida City, a mountainous area about 220\,km west of Tokyo. LISM showed the merit of going underground and demonstrated stable operation of the 20\,m interferometer. Kamioka is located in the Gifu Prefecture in a relatively quiet region of Japan, and the seismic noise level in the underground facility was extremely low (i.e. acceleration levels about 100 to 1000 times less than at the Mitaka campus for frequencies above 1\,Hz). Moreover interferometer controls were more robust due to the excellent temperature stability of about $0.01^\circ$C in the underground laboratory itself.

The LISM project was followed in 2006 by the CLIO (Cryogenic Laser Interferometer Observatory) project in which a 100\,m prototype interferometer started underground operation at Kamioka. Cooling of the mirrors to temperatures as low as 20\,K was pursued to reduce the thermal noise of both the mirrors and the lowest pendulum stage. Thermal noise is the major noise source in the signal band when seismic and quantum noises are sufficiently suppressed. Cryogenic cooling poses a variety of challenges because the cooling system creates vibrations. On the one hand, these vibrations should not reach the mirrors, but in order to keep the mirrors cold, there needs to be physical contact between the suspensions and the cold bath. CLIO had a Fabry-Perot interferometer configuration equipped with a ring mode cleaner for laser frequency stabilization and a cooling system using low vibration pulse-tube coolers to operate sapphire mirrors at 20\,K. 

In parallel with the upgrade projects Advanced LIGO and Advanced Virgo, Japanese scientists proposed the LCGT (Large-scale Cryogenic Gravitational-wave Telescope) project, a 3\,km interferometer with cryogenic mirrors to be constructed in the Ikenoyama mountain in Kamioka. In June 2010,
Japan's parliament approved funding for the project and in January 2012, it was given its new name, KAGRA (with ``KA" for Kamioka and ``GRA" for gravity). The name refers to \emph{kagura}, which is a type of Shinto ritual ceremonial dance. KAGRA is led by Takaaki Kajita (who was awarded the 2015 Nobel Prize in Physics for the discovery of neutrino oscillations) and managed by the ICRR of the University of Tokyo. Unique to the Japanese research funding system is the separation of human and material resources. This meant that the core team for setting up the interferometer was initially quite small given the magnitude of the project. Nevertheless, KAGRA made rapid progress in establishing the infrastructure and interferometer components. 

\begin{figure}[ht]
\includegraphics[width=\textwidth]{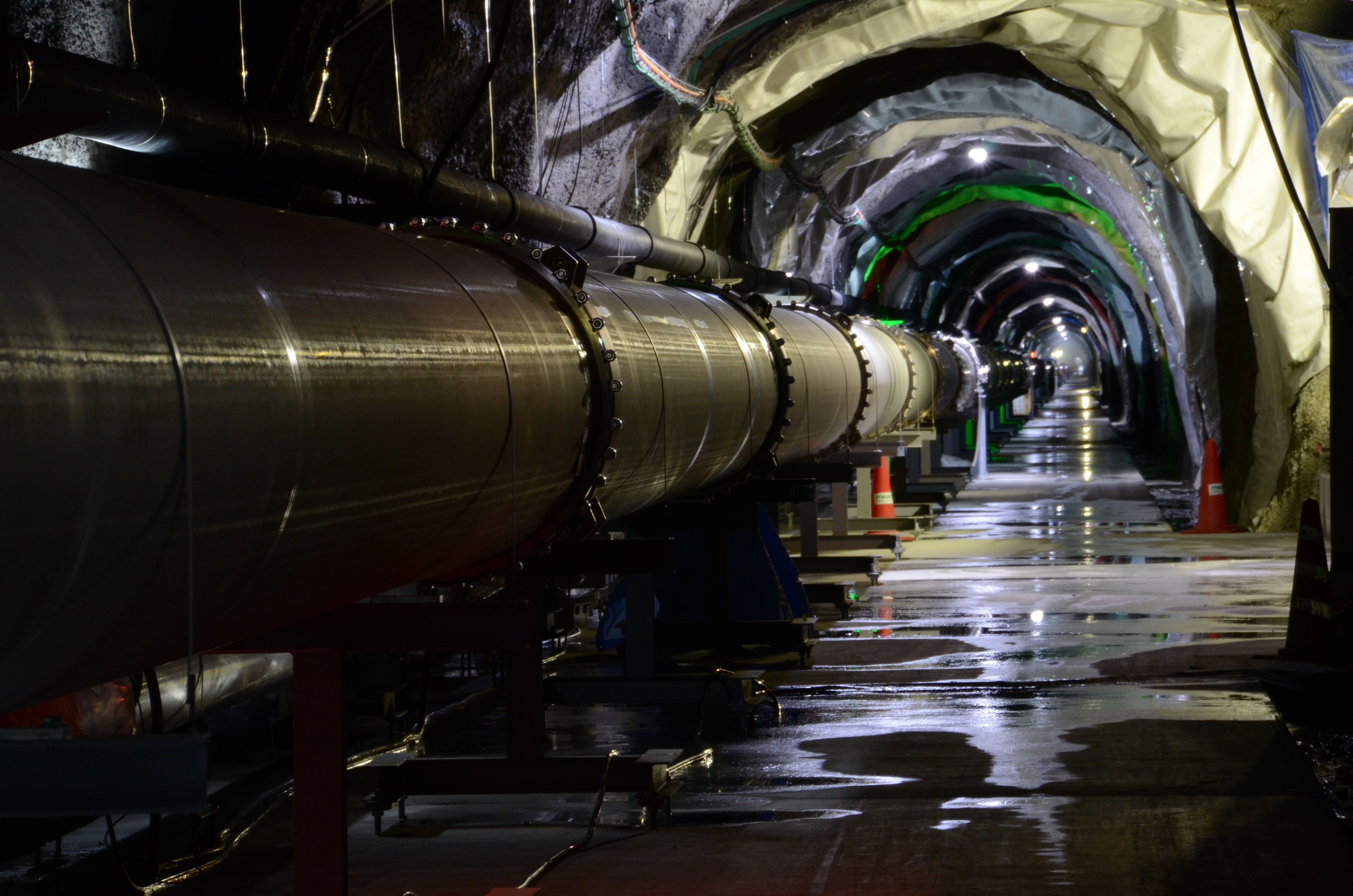}
\centering
\caption{View inside the KAGRA underground tunnel. The beam tube is 80\,cm in diameter and 3\,km long. Photo courtesy: KAGRA Observatory, ICRR, The University of Tokyo.}
\label{fig:KAGRA}       
\end{figure}

KAGRA will be a dual-recycled laser interferometric gravitational wave detector with Fabry-Perot cavities; see Fig. \ref{fig:KAGRA}. It is built in the same mountain with the neutrino physics experiments (Ikenoyana mountain). The construction of the tunnels was started in May 2012 and was completed in less than two years on March 31, 2014. The excavation of the two, three-kilometer-long tunnels for the arms was hampered due to severe water inrushes. Excess water in the tunnels caused significant delays in 2015. 

An initial interferometer consisting of single pendulum suspensions in a simple Michelson configuration was successfully brought to a brief operation phase in 2016. The installation of all optics for the full cryogenic interferometer with multiple-pendulum suspensions was nearly completed in early 2019 and commissioning of parts of the full interferometer had begun. KAGRA is the only interferometer that uses sapphire instead of silica mirrors and this presented new challenges due to birefringence of the mirror substrates. 

In October 2019 KAGRA signed a memorandum of agreement with LIGO and VIRGO to join the global network of gravitational wave detectors and on February 25, 2020, KAGRA joined Advanced LIGO and Advanced Virgo in the O3 observing run.
From April 7 to 21, 2020, KAGRA carried out a joint observation run (termed O3GK) with GEO\,600. While KAGRA's sensitivity is modest at present, it is expected to rapidly increase with time, as the performance of the instrumentation improves.

KAGRA \cite{KAGRA} is the world's first gravitational wave observatory in Asia, the first built underground, and the first detector that uses cryogenic mirrors. KAGRA pursues several of the key technologies that are needed for future detectors such as Einstein Telescope or Cosmic Explorer, and its experience will be of great importance to the global community. With KAGRA joining the network of advanced gravitational wave detectors, the location of sources will eventually be narrowed down to patches of sky that are only about 10 square degrees, enhancing the ability of light-based telescopes to carry out follow-up observations.

\subsection{GEO\,600}\index{GEO\,600}

Inspired by the progress and findings made on the prototypes in Munich and Garching, the German gravitational wave researchers tried to obtain funds in 1985 for a detector with an arm length of three kilometers. The German research funding organizations were not sufficiently interested, however, and in 1986 a similar situation occurred in the UK, where the application by Jim Hough's group for funding to build a large interferometer was also rejected. 

In the following years, these two groups merged to form the GEO Collaboration\index{GEO collaboration} and a German-British research proposal was submitted in 1989. In addition to experimental centres in Garching (Germany) and Glasgow (Scotland), groups from Cardiff University (Wales), University of Strathclyde (Glasgow), Max Planck Institute (Hanover, Germany), University of Oxford (UK) and Technical University of Braunschweig (Germany) participated. This application was submitted to the then Federal Ministry of Research and Technology in Germany and the British Science and Engineering Research Council. It proposed the construction of either an interferometer with an arm length of 3\,km in the German Harz mountains or a facility with an arm length of 2.6\,km in Scotland. The costs were estimated at an equivalent of about $\$50$ million. Despite positive evaluation, the project was ultimately rejected. In the aftermath of German reunification, in 1990, gravitational waves was not on the top of the list of German science policy priorities. 

After this disappointment, Karsten Danzmann, who had taken over the management of the Max Planck Group in Garching at the end of 1989, came up with the plan to build a much smaller, and therefore cheaper, facility. Thanks to ambitious technology, the German-British GEO\,600 detector was anticipated to be competitive with the larger detectors of the first generation in at least part of the frequency band. After the University of Hannover and the State of Lower Saxony provided suitable land in Ruthe near Sarstedt, south of Hannover, Germany, the GEO\,600 detector with an arm length of 600 meters was created. The German portion of the financing was contributed by the Max Planck Society, the Volkswagen Foundation and the State of Lower Saxony, and the British portion by the Particle Physics and Astronomy Research Council. The construction of GEO\,600 began in September 1995, with a large portion of the detector's infrastructure built by the scientists and students. An innovative design of the 600-m long vacuum tubes and a reduction to the bare essentials of the buildings made it possible to save money on material and infrastructure costs. 

The optical configuration of GEO\,600 consists of a Michelson interferometer\index{Michelson interferometer} with dual recycling\index{dual recycling}, the combination of power and signal recycling. Fabry-Perot resonators\index{Fabry-Perot resonator} were not used in the arms because the simultaneous use of arm resonators and dual recycling had not yet been tested on prototypes. Initial LIGO did not use this combination for the same reason. Instead of arm resonators, GEO\,600 used simple folded arms (delay lines\index{delay line} with just one additional mirror in each arm) to double the effective path length. From 2003 to 2009, GEO\,600 was operated in a moderately narrow-band mode of signal recycling\index{signal recycling}, which included the development of new techniques for mirror angle adjustments and for lock acquisition.

The use of ambitious technology of GEO\,600 proved a unique opportunity to test it for possible use in the larger detectors in the future.
From the beginning, GEO\,600 used triple pendulum\index{pendulum} suspensions and, since 2003, glass fibers for the suspensions of all test masses in the lowest pendulum stages. In comparison to conventional steel wires, glass fibers result in lower mechanical friction, which reduces thermal noise\index{thermal noise}. In addition, electrostatic actuators, which do not require that magnets be glued to the test masses, were the method chosen to exert forces on the test masses. The routine operation of these innovations at GEO\,600 bolstered confidence in these techniques. The triple pendulum suspension, including glass fibers and electrostatic actuators, were then further developed by the GEO collaboration into Advanced LIGO's four-stage suspension of test masses. 

The sensitivity of GEO\,600 did not reach that of LIGO, however, mainly due to the fact that LIGO's arms are seven times longer. 
The specific design choices of GEO\,600 also brought a challenge the other interferometers did not face to the same extent: Without Fabry-Perot resonators in the arms, all of the laser power must pass through the beam splitter. Inevitably, a small proportion of the light is absorbed by the beam splitter, heating it up and causing a lensing effect. 
The returning beams from both arms thus have slightly different shapes and can no longer fully destructively interfere. As a result, some light leaves the interferometer at the anti-symmetric port, giving rise to a variety of problems; power recycling may not work as well and there is more scattered light. Targeted heating of certain areas of the beam splitter or the test masses can reduce the effects of the beam splitter thermal lensing.

From 2002 to 2010, GEO\,600 participated in joint science runs with LIGO and Virgo. However, unlike LIGO and Virgo, no second generation interferometer was planned for GEO\,600. Instead, starting in 2008 an incremental upgrade program was implemented in which new techniques such as squeezed light application were tested. And from 2010 to 2015, while LIGO and Virgo were shut down for their respective upgrades, GEO\,600 was the only interferometer recording observational data in a program known as Astrowatch, which minimized the risk of missing a possible cosmic event of great strength. 
In the period from 2010 to 2018, GEO\,600 recorded observational data, on average, two thirds of the time. A high degree of automation allowed for such an intensive measurement operation and also made it possible for it to be carried out by a small team. 

In 2010, GEO\,600 had its most important upgrade\,---\,the application of squeezed vacuum\index{squeezed vacuum}, a technique that reduces shot noise \cite{GEOSqz:2011}. 
As early as 1981, the use of the squeezed vacuum technique to improve gravitational wave detectors was proposed by the American theoretical physicist Carlton Caves, but it took decades of laboratory work by groups in Germany, Australia and the US to make his idea a reality. 
While this new technology was briefly tested on a first-generation LIGO detector in 2011, it has been in permanent use at GEO\,600 since 2010 and is being continually improved. Due in large part to the demonstration and development of the technique at GEO\,600, Advanced LIGO and Advanced Virgo carried out early upgrades to install squeezed vacuum, which was implemented for the third observational run (O3) of 2019--2020. 

\section{Outlook}
\label{sec:outlook}
The global network of gravitational wave observatories now consists of the two Advanced LIGO instruments, Advanced Virgo, GEO\,600 and, recently, KAGRA. Moreover, the third Advanced LIGO interferometer that is under construction in India, known as LIGO-India, is expected to join the global network in 2027. We provide here a brief overview of ongoing research and plans for upgrades and new detectors. For more details we direct the reader to the dedicated chapters in this Handbook as well as a review article about the prospects for future observations \cite{OSP}.

\subsection{LIGO A+ and Virgo AdV+ upgrades}
The LIGO and Virgo collaborations aim to constantly improve the sensitivities of their instruments until the limits set by the sites are reached (e.g. the length of current interferometer arms cannot be extended significantly). The most recent upgrade projects are called A+ and AdV+ and involve improvements in instrumentation to the existing Advanced LIGO and Advanced Virgo instruments, respectively. Both A+ and AdV+ projects are underway with installation starting in 2020. A similar upgrade is under discussion for KAGRA.

\begin{itemize}
\item The \textbf{LIGO A+} upgrade will almost double Advanced LIGO's sensitivity, and increase the volume of space searched by a factor of up to seven. It includes improvements to the mirror suspension systems, including a larger beam splitter to reduce losses and increased reflectivities of the mirrors, in combination with new mirrors with improved coatings with lower mechanical loss. Furthermore, a $300$\,m long filter cavity will be installed in each interferometer to create frequency-dependent squeezed light, which facilitates simultaneously decreasing radiation pressure noise at low frequencies and shot noise at high frequencies. The gravitational-wave readout will use the balanced homodyne technique to reduce noise coupling from auxiliary control loops. The completion of the project is foreseen for 2024.

\vspace{2mm}
\item The \textbf{AdV+} upgrade will improve the sensitivity of Advanced Virgo by about a factor of two across a broad frequency range, with an expected completion of the project in 2024. AdV+ will be carried out in a phased approach. In Phase~1 (2020--2022) the sensitivity will be improved by using frequency-dependent squeezing and signal recycling. While signal recycling is already operational in LIGO, it still needs to be implemented and commissioned for Virgo. Frequency-dependent squeezing will reduce the fundamental quantum noise over the entire frequency range of the instrument. Sensitivity at low frequency will be improved through subtraction of gravity gradient noise. Phase~2 (2023--2025) will use larger beam sizes in combination with heavier mirrors (more than 100\,kg) and better coatings to overcome thermal noise. An increase of the beam power reduces shot noise, while an increase in the weight of the mirrors reduces the effects of radiation pressure. 

\end{itemize}

\subsection{LIGO-India}
\label{ssec:LIGO-India}
The third of the Advanced LIGO detectors will be located in India through an arrangement that was pursued by the LIGO Laboratory with support from NSF starting in 2011. A location in either the Southern hemisphere or in Asia for the third LIGO detector was preferable over the original plan of installing a second detector at the LIGO Hanford site in order to provide a longer baseline for reconstructing the sky location and the polarization of gravitational wave events \cite{Iyer:2011}. Detector hardware, including seismic isolation systems, mirrors, and electronics would be provided by the LIGO Laboratory and the site infrastructure by the partner host country.

Australia and India both expressed interest and ultimately LIGO-India was approved by the Indian government in 2016. Detailed surveys of 22 potential sites were conducted and the site at Aundha, near Hingoli in the state of Maharashtra was
identified as the most suitable location. Acquisition of the land was completed in 2019 and excavation and levelling work for the construction site office has begun as has the installation of fences. An off-site facility for a 10-m arm length prototype interferometer and testing bay for the seismic isolation system is nearing completion. Joint operation with Advanced LIGO, Advanced Virgo and KAGRA is expected by 2027.

\begin{figure}[ht]
\sidecaption
\centering
\includegraphics[width=\textwidth]{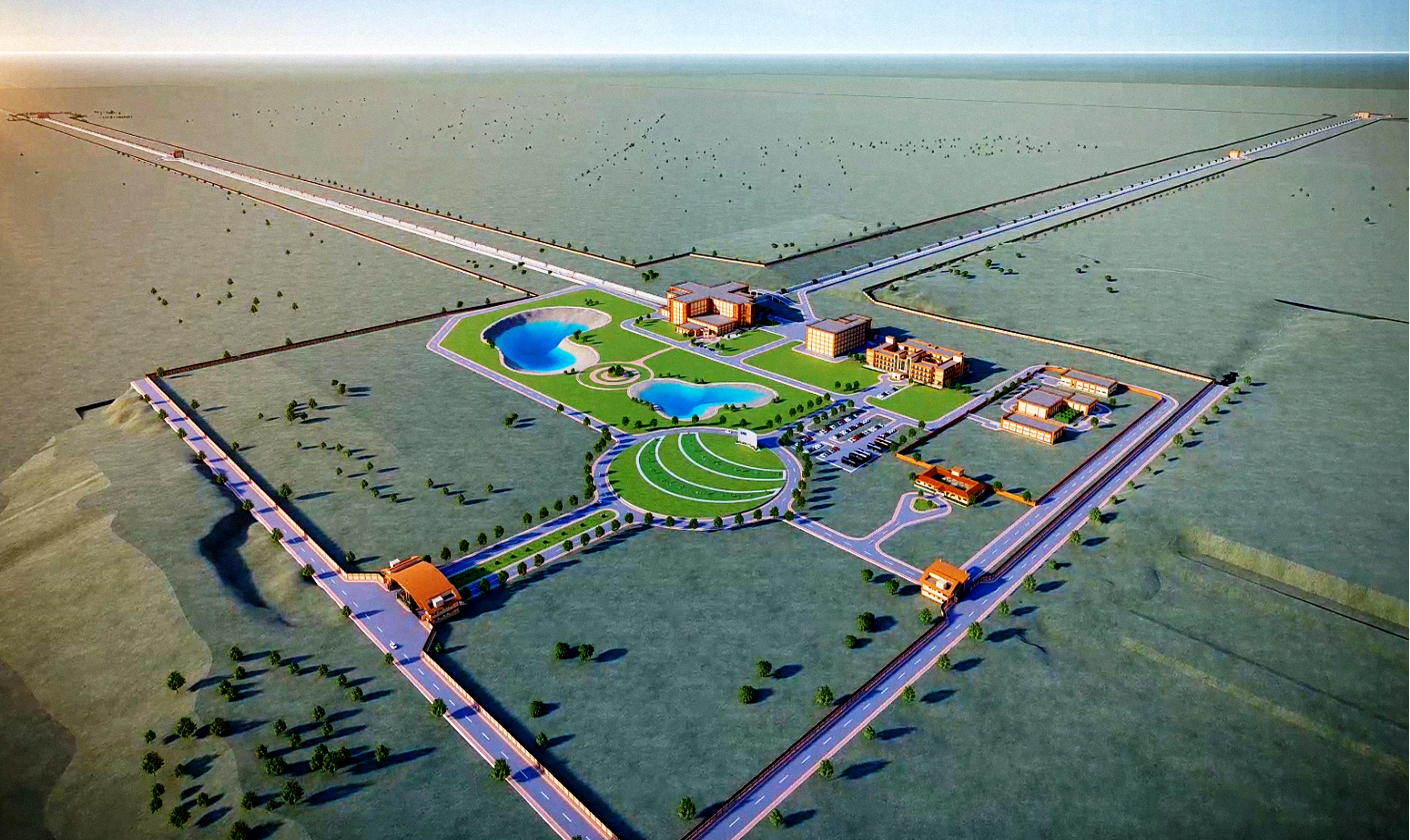}
\caption{Design concept of the LIGO-India Observatory based on actual architectural drawings and terrain data. Image courtesy: DCSEM-DAE/LI-EPO.}
\label{fig:LIGO-India}      
\end{figure}

\subsection{Third Generation Detectors}
In order to fully exploit the potential of gravitational wave science, new detectors in new facilities will eventually be required. These instruments would enable unprecedented cosmological observations, probe the densest regions of matter, the earliest stages after the Big Bang and the most extreme distortions of spacetime near black holes. These new observatories will annually detect up to a million gravitational wave transients from binary sources distributed throughout the entire Universe up to times as early as the ``dark ages", the era before the formation of the first stars. 
A phased approach is envisioned where more modest upgrades of the existing detectors would take place so that observing can continue while new facilities are built. Although funding is not yet secured, plans are well underway for the following so-called `3G' detectors:

\begin{itemize}
\item \textbf{LIGO Voyager} would be an upgrade to the LIGO A+ instruments at the existing sites, with the goal to improve the sensitivity by an additional factor of two. The silica mirrors in the current instruments would be replaced by larger and heavier (160\,kg) silicon test masses that are cooled to a temperature of 123\,K with liquid nitrogen. This requires the currently used 1064\,nm lasers to be replaced as silicon is only transparent between 1500 and 2200\,nm. Voyager would extend LIGO's sensitive frequency band to as low as 10\,Hz. The idea is for LIGO Voyager to be operational by 2030.

\vspace{2mm}
\item 
\textbf{Einstein Telescope} would be a future underground observatory in Europe. It will employ three detectors in a 10\,km triangular arrangement and a so-called xylophone configuration, where each detector consists of two distinct interferometers. One interferometer will be devoted to the detection of gravitational waves in the low $2$\,Hz to $40$\,Hz frequency range, while the other interferometer will focus on higher frequencies. The low-frequency interferometer will operate at cryogenic temperatures and the thermal, seismic, gravity gradient and radiation pressure noise sources will be particularly suppressed; the sensitivity of the high-frequency interferometer will be improved through the high laser power circulating in the Fabry-Perot cavities, and through the use of frequency-dependent squeezed light technologies. 

\smallskip
With the political support of five European countries, Belgium, Poland, Spain and The Netherlands, and led by Italy, a proposal to realize the Einstein Telescope infrastructure has been submitted to the 2021 update of the road map of the European Strategic Forum for Research. 
An Einstein Telescope consortium has been formed which consists of about 40 research institutions and universities located in several European countries, including also France, Germany, Hungary, Norway, Switzerland and the United Kingdom.

\vspace{2mm}
\item 
\textbf{Cosmic Explorer} is the name for a future facility in the US which is planned as a surface detector with a LIGO-type L-shape geometry but with 40\,km long arms.
Cosmic Explorer would ultimately be based on the LIGO Voyager technology, but an initial phase could be equipped
with the more conventional room temperature technology. 
Cosmic Explorer will have a higher sensitivity than Einstein Telescope for frequencies beyond 10\,Hz, but lower sensitivity below 10\,Hz. The design is under study and the Conceptual Design Report is anticipated for 2021.

\end{itemize}

\subsection{Final words}
It is remarkable to envision that in a time-span of only two decades the detection of gravitational waves may go from discovery of gravitational waves by the LIGO-Virgo Collaboration in 2015 to the realization of Einstein Telescope and Cosmic Explorer, innovative new infrastructures that will serve a worldwide scientific community, and that will ultimately lead to a better understanding of the origin and evolution of the Universe. This global network possesses great potential in contributing to fundamental physics, astrophysics, astronomy, cosmology and nuclear physics. 

\section*{Acknowledgments}
We thank Peter Saulson for many valuable comments on the manuscript, and we thank our colleagues in the LIGO, Virgo, GEO, and KAGRA collaborations for their marvelous work on the instruments and their components.

\section*{Cross-References}
\begin{itemize}
\item Introduction to gravitational wave astronomy
\item 3rd generation of GW detectors
\item R\&D for ground-based third-generation Gravitational wave interferometers
\item Squeezing and QM techniques in GW Interferometers
\item Environmental noise in ground based laser interferometers
\end{itemize}

\printindex
\bibliographystyle{ieeetr} 
\bibliography{references}

\end{document}